\documentclass[a4paper,fleqn,usenatbib,useAMS]{mnras}
\usepackage{graphicx}	
\usepackage{amsmath}	
\usepackage{amssymb}	
\usepackage{multicol}        
\usepackage{url}
\usepackage{soul} 
\usepackage{ulem} 
\usepackage{subfig}
\usepackage[T1]{fontenc}
\usepackage{ae,aecompl}
\usepackage{newtxtext,newtxmath}

\newcommand{\doublecolumnwidth}{480.0pt}
\newcommand{\doublecolumnwidthalt}{460.0pt}

\defcitealias{taylor09}{TSS09}

\title[Off-axis Instrumental Polarisation of the NVSS RM Catalogue]{A Broadband Spectro-polarimetric View of the NVSS Rotation Measure Catalogue II: Effects of Off-axis Instrumental Polarisation}

\author[Y.\ K.\ Ma et al.]{Yik Ki Ma$^1$\thanks{Contact e-mail: \href{mailto:ykma@mpifr-bonn.mpg.de}{ykma@mpifr-bonn.mpg.de}}\thanks{Member of the International Max Planck Research School (IMPRS) for Astronomy and Astrophysics at the Universities of Bonn and Cologne}, 
S.~A.~Mao$^1$, 
Jeroen Stil$^2$, 
Aritra Basu$^{3,1}$, 
Jennifer West$^4$, 
Carl Heiles$^5$, 
\newauthor 
Alex S.~Hill$^{6,7,8}$, and 
S.~K.~Betti$^9$
\\
$^1$Max-Planck-Institut f\"{u}r Radioastronomie, Auf dem H\"{u}gel 69, 53121 Bonn, Germany \\
$^2$Department of Physics and Astronomy, University of Calgary, 2500 University Drive NW, Calgary, AB T2N 1N4, Canada \\
$^3$Fakult\"{a}t f\"{u}r Physik, Universit\"{a}t Bielefeld, Postfach 100131, 33501 Bielefeld, Germany \\
$^4$Dunlap Institute for Astronomy and Astrophysics, The University of Toronto, 50 St.\ George Street, Toronto, ON M5S 3H4, Canada \\
$^5$Department of Astronomy, University of California, Berkeley, CA 94720-3411, USA \\
$^6$Department of Physics and Astronomy, The University of British Columbia, Vancouver, BC V6T 1Z1, Canada \\
$^7$Space Science Institute, Boulder, CO, USA \\
$^8$National Research Council Canada, Herzberg Program in Astronomy and Astrophysics, Dominion Radio Astrophysical Observatory, PO Box 248, Penticton, \\
\phantom{$^0$}BC V2A 6J9, Canada \\
$^9$Department of Astronomy, University of Massachusetts, 710 North Pleasant Street, Amherst, MA 01003-9305, USA
}

\date{Accepted 2019 May 10. Received 2019 May 7; in original form 2019 March 7}

\pubyear{2019}

\begin{document}
\label{firstpage}
\pagerange{\pageref{firstpage}--\pageref{lastpage}}
\maketitle

\begin{abstract}
The NRAO VLA Sky Survey (NVSS) Rotation Measure (RM) catalogue has enabled numerous studies in cosmic magnetism, and will continue being a unique dataset complementing future polarisation surveys. Robust comparisons with these new surveys will however require further understandings in the systematic effects present in the NVSS RM catalogue. In this paper, we make careful comparisons between our new on-axis broadband observations with the Karl G.\ Jansky Very Large Array and the NVSS RM results for 23 sources. We found that two unpolarised sources were reported as polarised at about 0.5\,\% level in the RM catalogue, and noted significant differences between our newly derived RM values and the catalogue values for the remaining 21 sources. These discrepancies are attributed to off-axis instrumental polarisation in the NVSS RM catalogue. By adopting the 0.5\,\% above as the typical off-axis instrumental polarisation amplitude, we quantified its effect on the reported RMs with a simulation, and found that on average the RM uncertainties in the catalogue have to be increased by $\approx 10\,\%$ to account for the off-axis instrumental polarisation effect. This effect is more substantial for sources with lower fractional polarisation, and is a function of the source's true RM. Moreover, the distribution of the resulting RM uncertainty is highly non-Gaussian. With the extra RM uncertainty incorporated, we found that the RM values from the two observations for most (18 out of 21) of our polarised targets can be reconciled. The remaining three are interpreted as showing hints of time variabilities in RM.
\end{abstract}


\section{INTRODUCTION} \label{sec:intro}
Magnetic fields are known to be crucial for astrophysical processes such as star formation, cosmic ray propagation, galactic outflows, and galactic evolution \cite[see, e.g.,][]{beck13,beck15}. While the magnetic field strength and structure of astrophysical objects can be probed by measurements of their polarised synchrotron emission \citep[e.g.,][]{fletcher11,giessuebel13,mao15,kierdorf17}, this method is only sensitive to the magnetic field component in the plane of the sky in volumes populated with cosmic ray electrons. A complementary method is to use background polarised sources as probes to the foreground subjects of interest -- polarised emission experiences the Faraday rotation effect as it traverses through the foreground intervening magnetised plasma, leading to a change in the polarisation position angle (PA; [rad]) given by
\begin{equation}
\Delta{\rm PA} = \left[ 0.81 \int_\ell^0 n_e(s) B_\parallel(s)\,{\rm d}s \right] \cdot \lambda^2 \equiv {\rm RM} \cdot \lambda^2{\rm ,}
\end{equation}
where $\ell$ [pc] is the (physical) distance to the source from the observer, $n_e$ [${\rm cm}^{-3}$] is the thermal electron density, $B_\parallel$ [$\mu$G] is the strength of the magnetic field component along the line of sight ($s$ [pc]), $\lambda$ [m] is the wavelength of the emission, and RM [${\rm rad\,m}^{-2}$] is the rotation measure of the source\footnote{In this work, we investigate the narrowband results presented in \cite{taylor09}, and therefore follow the traditional notion of RM instead of the more generalised notion of Faraday depth.}. The integrated value of the magnetic field strength along the line of sight, weighted by $n_e$, is therefore encrypted in the RM values. The RM of any given sightline can be obtained by PA measurements at two or more frequency bands, followed by a linear fit to PA against $\lambda^2$. For example, the resulting RM from observations at two frequencies only is given by
\begin{equation}
{\rm RM} = \frac{{\rm PA}_1 - {\rm PA}_2 + n\pi}{\lambda_1^2 - \lambda_2^2} \label{eq:rm},
\end{equation}
where the subscripts denote the two frequency bands, and $n$ is an integer corresponding to $n\pi$-ambiguity resulting from the possible wrapping(s) of PA between the two bands \citep[see Paper I,][for more details]{ma19a}.

Extragalactic radio sources (EGSs) have commonly been used to uncover the magnetic fields in foreground astrophysical objects. Such RM-grid experiments can be broadly divided into two categories: blind surveys and pointed observations. The NRAO VLA Sky Survey (NVSS) RM catalogue \citep[][hereafter \citetalias{taylor09}]{taylor09} is the largest RM catalogue to date, and is a notable example of blind surveys. This RM catalogue was built by re-analysing the original NVSS data \citep{condon98}, which were taken by scanning through a regularly spaced hexagonal grid in the northern sky ($\delta > -40^\circ$). The wide sky coverage of the \citetalias{taylor09} catalogue has enabled numerous studies of cosmic magnetism \citep[e.g.,][]{mcclure-griffiths10,harvey-smith11,stil11,oppermann12,hill13,oppermann15,purcell15,terral17}. Similar blind survey strategies were also adopted by other works for specific parts of the sky \citep[e.g.,][]{gaensler05,mao08,giessuebel13}. With such surveying strategies, the target EGSs are in general not on the pointing axis of the telescopes, which means the resulting data can be affected by off-axis instrumental effects that need to be accounted for (see below). On the other hand, the strategy of pointed observations is also commonly used \citep[e.g.,][]{mao10,vaneck11,mao12,costa16,kaczmarek17,mao17,betti19}, where target EGSs are selected from existing catalogues of polarised radio sources. They are then observed with the EGSs placed on the pointing axis of the telescopes. Compared to blind surveys, the resulting data of the targets from these pointed observations are free of off-axis instrumental artefacts.

An ideal radio telescope with dual polarised feeds should have independent polarisation channels, each being insensitive to its orthogonal counterpart. In reality, however, imperfections of the telescopes allow these polarisation channels to ``see'' the orthogonally polarised components. This is known as the instrumental polarisation (also known as the polarisation leakage) of radio telescopes, which can alter the measured polarisation signals. The polarisation leakage can be seen as comprised of two distinct elements -- the on-axis and the off-axis components. The former is routinely calibrated out in polarisation studies \citep[see, e.g.,][]{hales17}, usually by either (1) observing a known unpolarised calibrator and attributing the measured polarisation signals as the instrumental response of the telescope, or (2) observing a calibrator over a range of parallactic angles to simultaneously determine the astrophysical and instrumental polarisation, given that the telescope is driven by altitude-azimuthal (alt-az) mounts. Both these strategies will remove the polarisation leakage at the pointing centre where the calibrator has been placed at (down to, e.g., $\lesssim 0.02$ per cent in our new VLA observations; see Paper I), but residual polarisation leakage remains for positions within the primary beam away from the pointing axis (thus ``off-axis''). This off-axis instrumental polarisation can be determined by holography scans \citep[e.g.\ in the NVSS;][]{cotton94,condon98} and subsequently be calibrated out. Alternatively, the A-projection algorithm \citep{bhatnagar08,bhatnagar13} can be further developed to characterise and correct for the off-axis polarisation leakage \citep[see, e.g.,][]{jagannathan17,jagannathan18}. This full Mueller A-projection requires an adequate knowledge of the antenna optics, and can be applied during the imaging step of data reduction.

The off-axis polarisation leakage present in the NVSS data, if completely uncorrected, can be up to 2.5 per cent \citep{cotton94}. However, as calibrations for this off-axis leakage have been applied in the image domain, the residual leakage remaining in the data products of the original NVSS (namely, images and the source catalogue) is $\approx 0.3$ per cent \citep{condon98}. As \citetalias{taylor09} constructed their RM catalogue by re-analysing the NVSS visibility data, the calibration for the off-axis leakage was \emph{not} applied, though the mosaicking done to form their images could have smoothed out the off-axis leakage pattern with respect to the NVSS pointing centres. It is therefore likely that the reported RM values in the \citetalias{taylor09} catalogue have been affected by off-axis polarisation leakage, with its effect still remain unaccounted for.

It is crucial to fully understand the limits of this NVSS RM catalogue given its relevance. Although ongoing polarisation surveys such as Polarization Sky Survey of the Universe's Magnetism \citep[POSSUM;][]{gaensler10} in 1130--1430\,MHz and VLA Sky Survey \citep[VLASS;][]{myers14} in 2--4\,GHz will provide us with drastically higher RM densities than \citetalias{taylor09}, these two surveys either do not have exact sky or frequency coverage, and both cover different time domains, compared to \citetalias{taylor09}. This means the \citetalias{taylor09} catalogue will continue being a unique dataset depicting the magnetised Universe.

In our Paper I \citep{ma19a}, we have explored the $n\pi$-ambiguity problem in \citetalias{taylor09} and concluded that there are likely more than 50 $n\pi$-ambiguity sources (with erroneous RM by $\pm 652.9\,{\rm rad\,m}^{-2}$) out of the total 37,543 in the NVSS RM catalogue. In addition, we found two sources that were reported as $\approx 0.5$ per cent polarised in \citetalias{taylor09} but were unpolarised in our new broadband Karl G.\ Jansky Very Large Array (VLA) observations. We attributed this discrepancy in polarisation levels to the off-axis polarisation leakage in the NVSS data, which has motivated our study here. In this paper, we perform a rigorous comparison between our new data of 23 sources with the results from \citetalias{taylor09}. Our goal here is to identify and quantify systematic errors affecting RM measurements that were unaccounted for in the \citetalias{taylor09} catalogue. The observational setup and data reduction procedures are outlined in Section~\ref{sec:obs}, and the results are presented in Section~\ref{sec:results}. We discuss the discrepancies between our new results and \citetalias{taylor09} in Section~\ref{sec:discussion}. The effects of off-axis instrumental polarisation on RM measurements are quantified by simulations in Section~\ref{sec:sim}. Finally, we make concluding remarks on this work in Section~\ref{sec:conclusion}. In the forthcoming Paper III, we incorporate our knowledge from both Papers I and II to create a refined \citetalias{taylor09} catalogue for future studies of cosmic magnetism.

\begin{table*}
\caption{Comparison between new VLA and \citetalias{taylor09} results \label{table:cutout}}
\begin{tabular}{lcccccc}
\hline
\multicolumn{1}{c}{Source} & ${\rm RM}_{\rm VLA}$$^a$ & ${\rm RM}_{\rm Tcut}$ & $\Delta {\rm RM}$$^b$ & $|\Delta {\rm RM}|/\sigma$$^b$ & $\Delta S/S_{\rm 1.4\,GHz}$$^c$ & $\alpha_{\rm L}$ \\
\multicolumn{1}{c}{(NVSS)} & (${\rm rad\,m}^{-2}$) & (${\rm rad\,m}^{-2}$) & (${\rm rad\,m}^{-2}$) & & (\%) & \\
\hline
 J111857$+$123442$^\dagger$ & $+81.3 \pm 13.5$ & $+194.1 \pm 5.9$\phantom{0} & $-112.8 \pm 14.7$\phantom{0} & $7.67$ & $-13.9$$^d$ & $-0.232 \pm 0.003$\\
 J084701$-$233701 & $+353.2 \pm 9.2$\phantom{00} & $+462.5 \pm 15.8$ & $-109.3 \pm 18.3$\phantom{0} & $5.99$ & $+7.5$ & $-0.233 \pm 0.012$\\
 J170934$-$172853$^\dagger$ & $+111.2 \pm 4.7$\phantom{00} & $+193.7 \pm 14.7$ & $-82.4 \pm 15.4$ & $5.34$ & $+33.0$ & $-0.077 \pm 0.007$\\
 J224549$+$394122$^{\dagger\odot}$ & $-269.7 \pm 2.1$\phantom{00} & $-279.4 \pm 1.2$\phantom{0} & $+9.7 \pm 2.4$ & $4.12$ & $+7.0$ & $-0.988 \pm 0.008$\\
 J094808$-$344010$^\dagger$ & $+394.5 \pm 16.1$\phantom{0} & $+330.1 \pm 11.6$ & $+64.4 \pm 19.9$ & $3.24$ & $-28.0$ & $-0.373 \pm 0.009$\\
 J090015$-$281758 & $+350.1 \pm 2.9$\phantom{00} & $+335.7 \pm 4.1$\phantom{0} & $+14.4 \pm 5.0$\phantom{0} & $2.87$ & $+1.4$ & $-0.693 \pm 0.006$\\
 J190255$+$315942$^\dagger$ & $+261.0 \pm 5.3$\phantom{00} & $+243.3 \pm 3.8$\phantom{0} & $+17.7 \pm 6.5$\phantom{0} & $2.73$ & $ -9.0$ & $-0.320 \pm 0.003$\\
 J094750$-$371528$^\odot$ & $+359.2 \pm 21.9$\phantom{0} & $+293.1 \pm 23.4$ & $+66.1 \pm 32.0$ & $2.07$ & $-9.3$ & $-0.755 \pm 0.013$\\
 J220927$+$415834 & $-340.0 \pm 5.2$\phantom{00} & $-356.0 \pm 6.0$\phantom{0} & $+16.0 \pm 8.0$\phantom{0} & $2.01$ & $-5.3$ & $-0.964 \pm 0.006$\\
J022915$+$085125$^\dagger$ & $-307.8 \pm 102.3$ & $-129.3 \pm 8.3$\phantom{0} & $-178.5 \pm 102.7$ & $1.74$ & $+3.9$ & $-0.581 \pm 0.006$\\
J092410$-$290606$^{\star\star}$ & $+533.0 \pm 8.4$\phantom{00} & $+513.5 \pm 9.1$\phantom{0} & $+19.4 \pm 12.4$ & $1.57$ & $-4.6$ & $-0.955 \pm 0.006$\\
J154936$+$183500$^?$ & $-473.5 \pm 14.4$\phantom{0} & $-441.7 \pm 14.9$ & $-31.8 \pm 20.7$ & $1.54$ & $-2.4$ & $-0.797 \pm 0.003$\\
J091145$-$301305$^{\dagger\star\star}$ & $+237.6 \pm 5.5$\phantom{00} & $+226.3 \pm 6.0$\phantom{0} & $+11.2 \pm 8.1$\phantom{0} & $1.38$ & $-4.4$ & $-0.923 \pm 0.011$\\
J093349$-$302700 & $+345.7 \pm 8.0$\phantom{00} & $+331.5 \pm 10.0$ & $+14.2 \pm 12.6$ & $1.12$ & $-9.0$ & $-0.972 \pm 0.007$\\
J220205$+$394913 & $-358.7 \pm 5.8$\phantom{00} & $-365.3 \pm 6.7$\phantom{0} & $+6.6 \pm 8.9$ & $0.75$ & $-2.3$ & $-1.140 \pm 0.009$\\
J093544$-$322845$^{\star\star}$ & $+440.5 \pm 25.8$\phantom{0} & $+411.0 \pm 33.2$ & $+29.6 \pm 42.0$ & $0.70$ & $-4.3$ & $-0.857 \pm 0.006$\\
J162706$-$091705$^{\star\star}$ & $-328.7 \pm 9.8$\phantom{00} & $-318.4 \pm 15.9$ & $-10.4 \pm 18.7$ & $0.55$ & $+0.9$ & $-1.061 \pm 0.013$\\
J235728$+$230226$^\dagger$ & \phantom{0}$+30.8 \pm 103.1$ & \phantom{0}$+82.0 \pm 13.1$ & \phantom{0}$-51.2 \pm 103.9$ & $0.49$ & $-0.6$ & $-0.863 \pm 0.007$\\
J083930$-$240723 & $+345.6 \pm 8.2$\phantom{00} & $+351.7 \pm 13.1$ & \phantom{0}$-6.1 \pm 15.5$ & $0.39$ & $-5.2$ & $-0.904 \pm 0.010$\\
J163927$-$124139$^{\star\star}$ & $-329.1 \pm 4.7$\phantom{00} & $-325.0 \pm 9.6$\phantom{0} & \phantom{0}$-4.1 \pm 10.7$ & $0.38$ & $+0.5$ & $-0.822 \pm 0.005$\\
J224412$+$405715$^\dagger$ & $-325.9 \pm 6.0$\phantom{00} & $-325.7 \pm 13.0$ & \phantom{0}$-0.2 \pm 14.3$ & $0.01$ & $+11.7$ & $+0.015 \pm 0.008$\\
J084600$-$261054$^\times$ & --- & --- & --- & --- & $-2.9$ & $-0.437 \pm 0.007$\\
J234033$+$133300$^\times$ & --- & --- & --- & --- & $+2.8$ & $-1.252 \pm 0.005$\\
\hline
\multicolumn{7}{l}{\texttt{NOTE}---Sorted by $|\Delta{\rm RM}|/\sigma$ in descending order} \\
\multicolumn{7}{l}{$^a$Using our new data in the NVSS bands only} \\
\multicolumn{7}{l}{$^b$$\Delta{\rm RM} = {\rm RM}_{\rm VLA} - {\rm RM}_{\rm Tcut}$} \\
\multicolumn{7}{l}{$^c$$\Delta S = S_{\rm 1.4\,GHz} - S_{\rm Ncut}$} \\
\multicolumn{7}{l}{$^d$$S_{\rm Ncut}$ from the original NVSS has been replaced by $S_{\rm Tcut}$ from \citetalias{taylor09} instead (see Section~\ref{sec:ivar})} \\
\multicolumn{7}{l}{$^\dagger$Out-\textit{liars} ($n\pi$-ambiguity sources; see Paper I)} \\
\multicolumn{7}{l}{$^\times$Unpolarised sources (less than the $6\sigma$ cutoff level)} \\
\multicolumn{7}{l}{$^?$Special case compared to \citetalias{taylor09} catalogue (see Paper I)} \\
\multicolumn{7}{l}{$^{\star\star}$Double point sources} \\
\multicolumn{7}{l}{$^\odot$Extended sources}
\end{tabular}
\end{table*}

\section{OBSERVATIONS AND DATA REDUCTION} \label{sec:obs}
\subsection{New Observations}
A total of 23 target sources were selected from the \citetalias{taylor09} catalogue, with the original primary goal of identifying $n\pi$-ambiguity sources (addressed in Paper I). These sources were selected based on their high $|{\rm RM}_{\rm TSS09}| \gtrsim 300\,{\rm rad\,m}^{-2}$, despite being situated away from the Galactic plane ($|b| > 10^\circ$)\footnote{Except for J234033$+$133300, which has ${\rm RM}_{\rm TSS09} = +56.7 \pm 6.3\,{\rm rad\,m}^{-2}$. This included source turned out to be unpolarised in our new observations, and is pivotal to our study of off-axis instrumental polarisation in this paper.}. Furthermore, all of our targets are bright with NVSS total intensities larger than 100\,mJy. 

\begin{figure}
\includegraphics[width=\columnwidth]{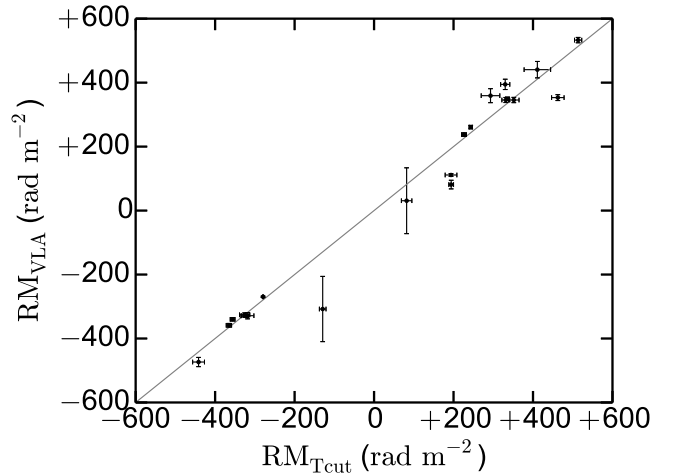}
\caption{Narrowband RM from our observations against that from \citetalias{taylor09} cutout images. The $n\pi$-ambiguities have been corrected by using our broadband $\overline\phi$ values from RM-Synthesis. The solid line shows where the RM values of the two measurements are identical. \label{fig:narrowrm}}
\end{figure}

Our new broadband observations were performed using the VLA in L-band (1--2\,GHz) in D array configuration on 2014 July 03. The same array configuration as used by the NVSS means that our \textit{uv}-coverages are similar to that of \citetalias{taylor09}, and the observations from both works are sensitive to emission at the same ranges of angular scales. The typical integration time per source was about 3--4 minutes. Standard calibration procedures were followed using the Common Astronomy Software Applications (CASA) package \citep[version 4.4.0;][]{mcmullin07}, and are described in detail in Paper I.

\subsection{NVSS Band Images from New Observations} \label{sec:nvssbandreduction}
A careful comparison in polarisation properties between our new data and the NVSS RM catalogue requires that the two datasets have near identical frequency and \textit{uv}-coverages, with the source properties extracted following the same method. We have therefore formed two sets of radio images using our calibrated broadband VLA data in the two NVSS intermediate frequency (IF) bands only. Although the original NVSS bands had frequency ranges of $1343.9$--$1385.9\,{\rm MHz}$ (IF1) and $1414.1$--$1456.1\,{\rm MHz}$ (IF2) respectively \citep{condon98}, parts of our data in these frequency ranges were unfortunately flagged because of corruption by radio frequency interferences (RFI) or because they lie at the edges of the new broadband VLA spectral windows where sensitivity drops rapidly. Therefore, we have instead used our new VLA data within frequency ranges of $1344.5$--$1373.5\,{\rm MHz}$ as IF1 and $1430.5$--$1445.5\,{\rm MHz}$ as IF2 uniformly for all the sources. These frequency ranges are the widest that we can get within the NVSS frequency bands, resulting in the best comparison that can be made between our data and \citetalias{taylor09} results.

Using our new VLA data, we created Stokes \textit{I}, \textit{Q}, and \textit{U} images for each source in the NVSS IF1 and IF2 respectively. All the images were made with Briggs visibilities weighting of \texttt{robust}$ = 0$ \citep{briggs95} with a common restoring beam of $60^{\prime\prime} \times 60^{\prime\prime}$ matching that of the \citetalias{taylor09} images (see below). On the other hand, we obtained cutout images of our target sources from \citetalias{taylor09}. The \citetalias{taylor09} images were formed using the calibrated NVSS visibility data, independently for NVSS IF1 and IF2. A mild \textit{uv}-taper was applied and led to their resulting beam of $60^{\prime\prime} \times 60^{\prime\prime}$. We decided to determine the RM values from the \citetalias{taylor09} images instead of directly adopting the listed ${\rm RM}_{\rm TSS09}$ values in their catalogue to ensure that the most direct comparison is performed. The Stokes \textit{I}, \textit{Q}, and \textit{U} values of our sources were extracted from the two observations in identical ways, with the flux densities of the unresolved sources determined from the CASA task \texttt{IMFIT}, and that of double and extended sources by integrating within $6\sigma$ contours in total intensity.

\begin{table*}
\caption{Total flux densities and redshifts of the targets \label{table:stokesi}}
\begin{tabular}{lcccccc}
\hline
\multicolumn{1}{c}{Source} & $\alpha_{\rm L}$ & $S_{1.4\,{\rm GHz}}$ & $S_{\rm Ncut}$$^a$ & $S_{\rm NVSS}$$^b$ & $z$ & Reference \\
\multicolumn{1}{c}{(NVSS)} & ($S_\nu \propto \nu^{\alpha_{\rm L}}$) & (mJy) & (mJy) & (mJy) & & (z) \\
\hline
J022915$+$085125 & $-0.581 \pm 0.006$ & \phantom{0}$649.2 \pm 0.8$ & $624.1 \pm 2.4$ & $609.0 \pm 18.3$ & --- & ---\\
J083930$-$240723 & $-0.904 \pm 0.010$ & \phantom{0}$261.6 \pm 0.4$ & $275.3 \pm 1.9$ & $268.3 \pm 9.3$\phantom{0} & --- & ---\\
J084600$-$261054 & $-0.437 \pm 0.007$ & $1787.6 \pm 2.0$ & $1839.9 \pm 2.1$ & $1810.6 \pm 54.3$\phantom{0} & --- & ---\\
J084701$-$233701 & $-0.233 \pm 0.012$ & \phantom{0}$162.1 \pm 0.3$ & $150.0 \pm 1.9$ & $145.0 \pm 4.4$\phantom{0} & $0.0607 \pm 0.0001$$^s$ & \cite{2mass}\\
J090015$-$281758 & $-0.693 \pm 0.006$ & \phantom{0}$530.8 \pm 0.6$ & $523.6 \pm 2.6$ & $511.6 \pm 15.4$ & $0.894$$^s$ & \cite{cxrb}\\
J091145$-$301305$^{\star\star}$ & $-0.923 \pm 0.011$ & \phantom{0}$238.6 \pm 0.5$ & $249.2 \pm 1.6$ & $247.1 \pm 7.8$\phantom{0} & --- & ---\\
\multicolumn{1}{c}{$\cdots$ a} & $-1.054 \pm 0.022$ & \phantom{00}$81.0 \pm 0.4$ & --- & --- & --- & ---\\
\multicolumn{1}{c}{$\cdots$ b} & $-0.858 \pm 0.010$ & \phantom{0}$157.6 \pm 0.3$ & --- & --- & --- & ---\\
J092410$-$290606$^{\star\star}$ & $-0.955 \pm 0.006$ & \phantom{0}$590.2 \pm 0.5$ & $617.4 \pm 1.9$ & $625.4 \pm 19.5$ & --- & ---\\
\multicolumn{1}{c}{$\cdots$ a} & $-1.011 \pm 0.010$ & \phantom{0}$291.0 \pm 0.5$ & --- & --- & --- & ---\\
\multicolumn{1}{c}{$\cdots$ b} & $-0.901 \pm 0.009$ & \phantom{0}$299.2 \pm 0.4$ & --- & --- & --- & ---\\
J093349$-$302700 & $-0.972 \pm 0.007$ & \phantom{0}$253.6 \pm 0.3$ & $276.5 \pm 2.1$ & $273.0 \pm 9.2$\phantom{0} & --- & ---\\
J093544$-$322845$^{\star\star}$ & $-0.857 \pm 0.006$ & \phantom{0}$499.3 \pm 0.7$ & $521.0 \pm 2.0$ & $244.6 \pm 7.4$\phantom{0} & --- & ---\\
\multicolumn{1}{c}{$\cdots$ a} & $-0.761 \pm 0.010$ & \phantom{0}$227.5 \pm 0.5$ & --- & --- & --- & ---\\
\multicolumn{1}{c}{$\cdots$ b} & $-0.941 \pm 0.008$ & \phantom{0}$271.9 \pm 0.5$ & --- & --- & --- & ---\\
J094750$-$371528$^\odot$ & $-0.755 \pm 0.013$ & \phantom{0}$602.3 \pm 1.6$ & $658.4 \pm 2.0$ & $473.9 \pm 15.1$ & $0.0412 \pm 0.0002$$^s$ & \cite{6dFGS}\\
J094808$-$344010 & $-0.373 \pm 0.009$ & \phantom{0}$245.6 \pm 0.4$ & $314.3 \pm 2.2$ & $312.4 \pm 9.4$\phantom{0} & --- & ---\\
J111857$+$123442 & $-0.232 \pm 0.003$ & $2041.5 \pm 1.0$ & $1129.4 \pm 0.9$$^c$ & $1112.2 \pm 33.4$$^c$ & \phantom{0}$2.125 \pm 0.0003$$^s$ & \cite{sdssdr14}\\
J154936$+$183500 & $-0.797 \pm 0.003$ & \phantom{0}$583.4 \pm 0.4$ & $597.4 \pm 2.4$ & $584.6 \pm 20.6$ & $1.442$$^s$ & \cite{hewitt87}\\
J162706$-$091705$^{\star\star}$ & $-1.061 \pm 0.013$ & \phantom{0}$130.8 \pm 0.4$ & $129.6 \pm 1.6$ & $125.0 \pm 4.3$\phantom{0} & --- & ---\\
\multicolumn{1}{c}{$\cdots$ a} & $-0.730 \pm 0.042$ & \phantom{00}$30.7 \pm 0.3$ & --- & --- & --- & ---\\
\multicolumn{1}{c}{$\cdots$ b} & $-1.171 \pm 0.015$ & \phantom{0}$100.2 \pm 0.3$ & --- & --- & --- & ---\\
J163927$-$124139$^{\star\star}$ & $-0.822 \pm 0.005$ & \phantom{0}$394.1 \pm 0.4$ & $392.0 \pm 2.3$ & $397.3 \pm 12.5$ & --- & ---\\
\multicolumn{1}{c}{$\cdots$ a} & $-0.810 \pm 0.006$ & \phantom{0}$247.5 \pm 0.3$ & --- & --- & --- & ---\\
\multicolumn{1}{c}{$\cdots$ b} & $-0.843 \pm 0.010$ & \phantom{0}$146.6 \pm 0.3$ & --- & --- & --- & ---\\
J170934$-$172853 & $-0.077 \pm 0.007$ & \phantom{0}$644.9 \pm 0.9$ & $432.2 \pm 2.5$ & $431.1 \pm 12.9$ & --- & ---\\
J190255$+$315942 & $-0.320 \pm 0.003$ & $2923.5 \pm 2.0$ & $3186.9 \pm 2.9$ & $3203.8 \pm 96.1$\phantom{0} & $0.635$$^s$ & \cite{hewitt87}\\
J220205$+$394913 & $-1.140 \pm 0.009$ & \phantom{0}$143.8 \pm 0.2$ & $147.1 \pm 1.7$ & $144.8 \pm 5.2$\phantom{0} & --- & ---\\
J220927$+$415834 & $-0.964 \pm 0.006$ & \phantom{0}$283.5 \pm 0.3$ & $298.6 \pm 1.7$ & $292.4 \pm 9.8$\phantom{0} & $0.521 \pm 0.029$$^p$ & \cite{sdssdr14}\\
J224412$+$405715 & $+0.015 \pm 0.008$ & \phantom{0}$260.4 \pm 0.4$ & $229.9 \pm 2.4$ & $226.2 \pm 6.8$\phantom{0} & $1.171$$^s$ & \cite{fermiagn2}\\
J224549$+$394122$^\odot$ & $-0.988 \pm 0.008$ & $11181.1 \pm 16.0$ & $10393.7 \pm 5.5$ & $4408.1 \pm 136.4$ & 0.081$^s$ & \cite{hewitt91}\\
J234033$+$133300 & $-1.252 \pm 0.005$ & $1868.0 \pm 1.7$ & $1815.9 \pm 2.3$ & $1829.1 \pm 54.9$\phantom{0} & --- & ---\\
J235728$+$230226 & $-0.863 \pm 0.007$ & \phantom{0}$634.9 \pm 0.9$ & $638.6 \pm 2.3$ & $624.7 \pm 18.7$ & $0.420 \pm 0.120$$^p$ & \cite{sdssdr14} \\
\hline
\multicolumn{7}{l}{$^a$Our integrated flux densities from NVSS cutout images} \\
\multicolumn{7}{l}{$^b$Listed integrated flux densities from the NVSS catalogue \citep{condon98}} \\
\multicolumn{7}{l}{$^c$Likely erroneous due to missing pointing in NVSS (see Section~\ref{sec:ivar})} \\
\multicolumn{7}{l}{$^{\star\star}$Double point sources} \\
\multicolumn{7}{l}{$^\odot$Extended sources} \\
\multicolumn{7}{l}{$^p$Photometric redshifts} \\
\multicolumn{7}{l}{$^s$Spectroscopic redshifts}
\end{tabular}
\end{table*}

\begin{table}
\caption{Positions of individual components of the spatial doubles \label{table:double}}
\begin{tabular}{ccc}
\hline
Source & Right Ascension & Declination \\
(NVSS) & (J2000; h m s) & (J2000; $^\circ$ $^\prime$ $^{\prime\prime}$) \\
\hline
J091145$-$301305 && \\
\multicolumn{1}{c}{$\cdots$ a} & 09 11 42.47 $\pm$ 0.04 & $-$30 13 19.26 $\pm$ 1.45 \\
\multicolumn{1}{c}{$\cdots$ b} & 09 11 46.33 $\pm$ 0.02 & $-$30 12 58.63 $\pm$ 0.74 \\
J092410$-$290606 && \\
\multicolumn{1}{c}{$\cdots$ a} & 09 24 10.09 $\pm$ 0.02 & $-$29 05 45.36 $\pm$ 0.79 \\
\multicolumn{1}{c}{$\cdots$ b} & 09 24 11.44 $\pm$ 0.02 & $-$29 06 26.66 $\pm$ 0.75 \\
J093544$-$322845 && \\
\multicolumn{1}{c}{$\cdots$ a} & 09 35 43.98 $\pm$ 0.02 & $-$32 28 48.51 $\pm$ 0.65 \\
\multicolumn{1}{c}{$\cdots$ b} & 09 35 43.79 $\pm$ 0.02 & $-$32 29 40.03 $\pm$ 0.60 \\
J162706$-$091705 && \\ 
\multicolumn{1}{c}{$\cdots$ a} & 16 27 04.53 $\pm$ 0.02 & $-$09 16 55.99 $\pm$ 0.64 \\
\multicolumn{1}{c}{$\cdots$ b} & 16 27 06.78 $\pm$ 0.01 & $-$09 17 06.50 $\pm$ 0.20 \\
J163927$-$124139 && \\
\multicolumn{1}{c}{$\cdots$ a} & 16 39 27.09 $\pm$ 0.01 & $-$12 41 26.41 $\pm$ 0.15 \\
\multicolumn{1}{c}{$\cdots$ b} & 16 39 28.20 $\pm$ 0.01 & $-$12 42 09.07 $\pm$ 0.27 \\
\hline
\multicolumn{3}{l}{\texttt{NOTE} -- Identical to Table 2 of Paper I}
\end{tabular}
\end{table}

\section{RESULTS}\label{sec:results}
\subsection{Rotation Measure Comparison with \citetalias{taylor09} \label{sec:nvss_comp}}
We perform a rigorous comparison between the polarisation properties of our new VLA results and that in \citetalias{taylor09}. By using the Stokes \textit{Q} and \textit{U} values obtained from our VLA data in NVSS bands and \citetalias{taylor09} cutout images, we computed the PA values by
\begin{equation}
{\rm PA}_j = \frac{1}{2} \tan^{-1} \left( \frac{U_j}{Q_j} \right){\rm ,}\label{eq:2band_rm}
\end{equation}
where the subscripts denote the IFs. Equation~\ref{eq:rm} is then used to compute the RM values. We used $\lambda_1 = 0.2196\,{\rm m}$ and $\lambda_2 = 0.2089\,{\rm m}$ for \citetalias{taylor09} cutouts, and $\lambda_1 = 0.2206\,{\rm m}$ and $\lambda_2 = 0.2085\,{\rm m}$ for our VLA data. The values of $n$ were chosen such that the RM values would most closely match the broadband polarisation-weighted Faraday depths ($\overline\phi$) reported in Paper I, which is free of $n\pi$-ambiguity. The only exception is J154936$+$183500, since we did not find correspondence between our broadband $\overline\phi$ and the narrowband ${\rm RM}_{\rm TSS09}$ for this source\footnote{We determined in our Paper I that the disagreement between $\overline\phi$ and ${\rm RM}_{\rm TSS09}$ for this source is because of its highly non-linear PA across $\lambda^2$ resulting from its significant Faraday complexities.}, and therefore we chose $n$ for this source such that the resulting RM values are the closest to its \citetalias{taylor09} value of $-426.8\,{\rm rad\,m}^{-2}$. As a sanity check, we further compared our RM values from \citetalias{taylor09} cutout images with the officially listed ${\rm RM}_{\rm TSS09}$ (corrected for $n\pi$-ambiguity as above) to ensure that these two sets of RM values agree with each other (within $\approx 1\sigma$), as would be expected since they were computed from the same dataset.

The RM values obtained in the two datasets are shown in Figure~\ref{fig:narrowrm}, and listed in Table~\ref{table:cutout}. The RM values from our new observations are denoted as ${\rm RM}_{\rm VLA}$, and that from \citetalias{taylor09} cutout images as ${\rm RM}_{\rm Tcut}$. We also listed the difference in RM ($\Delta{\rm RM} = {\rm RM}_{\rm VLA} - {\rm RM}_{\rm Tcut}$), as well as its magnitude divided by RM uncertainties (i.e., RM differences in units of $\sigma$; $|\Delta{\rm RM}|/\sigma$) in the same Table~\ref{table:cutout}. It is found that nine (43 per cent) and five (24 per cent) out of our 21 polarised sources have deviating RM values by more than $2$ and $3\sigma$ respectively. Assuming that the RM uncertainties are Gaussian (which is approximately true given the high signal-to-noise ratio in polarisation of $\gtrsim 15$ for these sources), $|\Delta{\rm RM}|/\sigma$ should follow a folded normal distribution with $\mu = 0$ and $\sigma = 1$ (corresponding to the mean and standard deviation of the parent normal distribution, respectively). This distribution predicts much lower, namely, 4.6 and 0.3 per cent of the data deviating by more than $2$ and $3\sigma$ respectively. It is evident that, with the entire sample of 21 sources considered as a whole, the RM values from \citetalias{taylor09} and our observations do not agree within their uncertainties.

\subsection{Flux Densities and Spectral Indices}
We report here the full-band (1--2\,GHz) radio spectra of our targets. This has been deferred to this paper from Paper I because of the relevance between Stokes \textit{I} and RM time variabilities. The flux densities were extracted from 4\,MHz channel images using the entire L-band (see Paper I), and were fitted for each source with a simple power law:
\begin{equation}
S_\nu = S_{1.4\,{\rm GHz}} \cdot \left( \frac{\nu}{1.4\,{\rm GHz}} \right)^{\alpha_{\rm L}}{\rm ,}\label{eq:pl}
\end{equation}
where $S_\nu$ is the flux density at frequency $\nu$, and $\alpha_{\rm L}$ is the spectral index in L-band. The values of $S_{1.4\,{\rm GHz}}$ and $\alpha_{\rm L}$ are listed in Table~\ref{table:stokesi}, with the radio spectra shown in Figure~\ref{fig:stokesi}. For double sources, apart from fitting the radio spectra of each of the spatial components (listed in Table~\ref{table:double}), we also added the flux densities together (with uncertainties added in quadrature) to obtain a joint fit, which facilitates comparison with the integrated flux densities reported in the NVSS ($S_{\rm NVSS}$; also in Table~\ref{table:stokesi}), as well as other lower resolution radio studies. We do not see clear evidence of deviations of the radio spectra from the simple power law for any of the sources.  We note the significant discrepancies between $S_{\rm 1.4\,GHz}$ and $S_{\rm NVSS}$ for the double / extended sources J093544$-$322845, J094750$-$371528, and J224549$+$394122. The differences may be due to sub-optimal spatial fitting to these resolved sources in the NVSS catalogue. We therefore extracted the flux densities of our sample from NVSS cutout images ($S_{\rm Ncut}$), also listed in Table~\ref{table:stokesi}. Indeed, we see much better agreement between $S_{\rm 1.4\,GHz}$ and $S_{\rm Ncut}$ for the above three sources, but for four of our other targets (J094808$-$344010, J111857$+$123442, J170934$-$172853, and J224412$+$405715) we still see differences in flux densities by more than 10 per cent (see Section~\ref{sec:ivar}).

\section{Comparisons with NVSS Results} \label{sec:discussion}

\begin{figure*}
\includegraphics[width=\doublecolumnwidthalt]{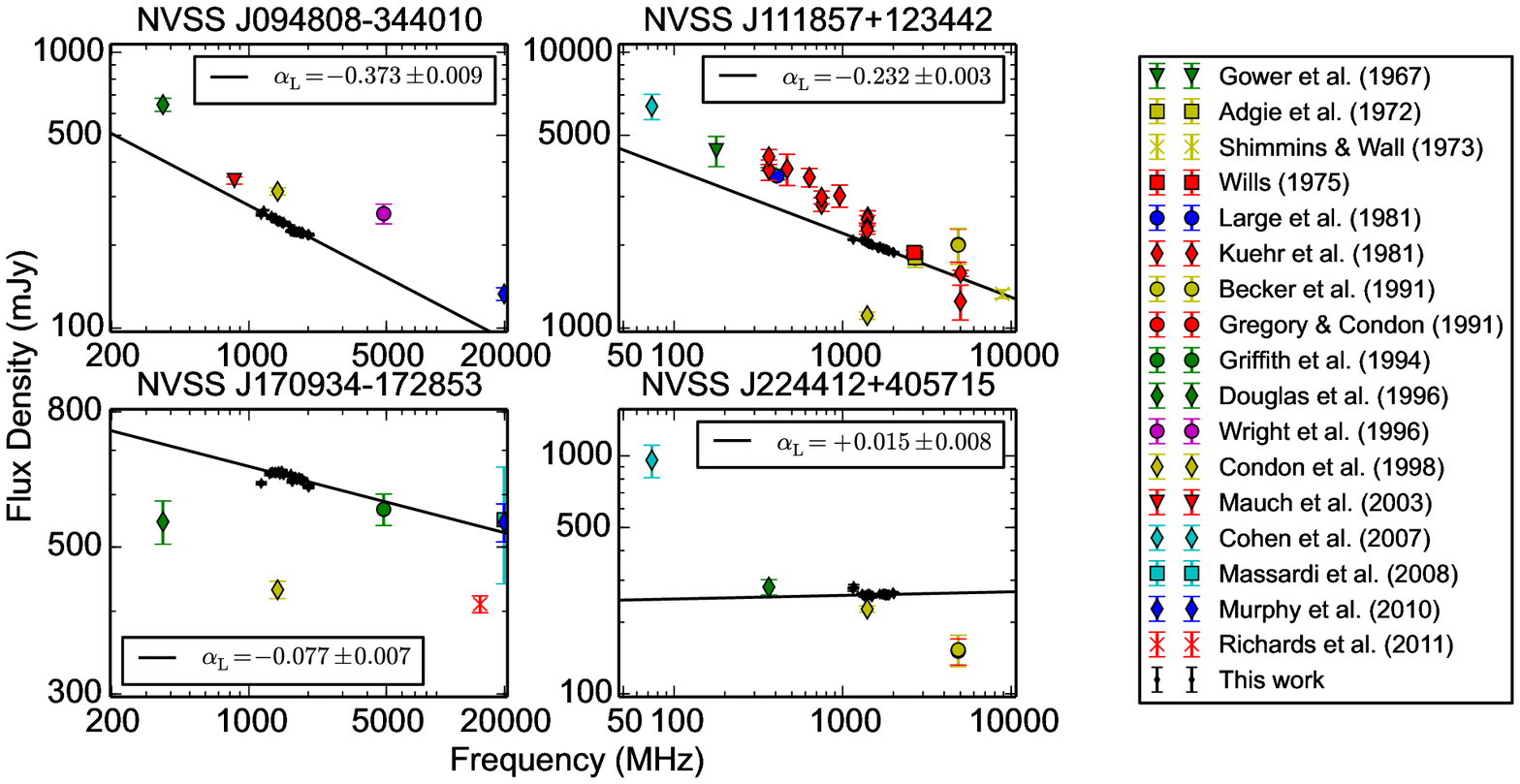}
\caption{Radio spectra of the four sources with signs of Stokes \textit{I} variabilities. Flux densities obtained from our new broadband measurements are shown as the black points, with the black solid line representing the best-fit power law spectra. The coloured data points represent flux densities from various studies. \label{fig:stokesi_all}}
\nocite{murphy10,wright96,mauch03,douglas96,shimmins73,kuehr81,gregory91,becker91,wills75,adgie72,large81,gower67,cohen07,massardi08,richards11,griffith94}
\end{figure*}

\begin{figure*}
\includegraphics[width=\doublecolumnwidthalt]{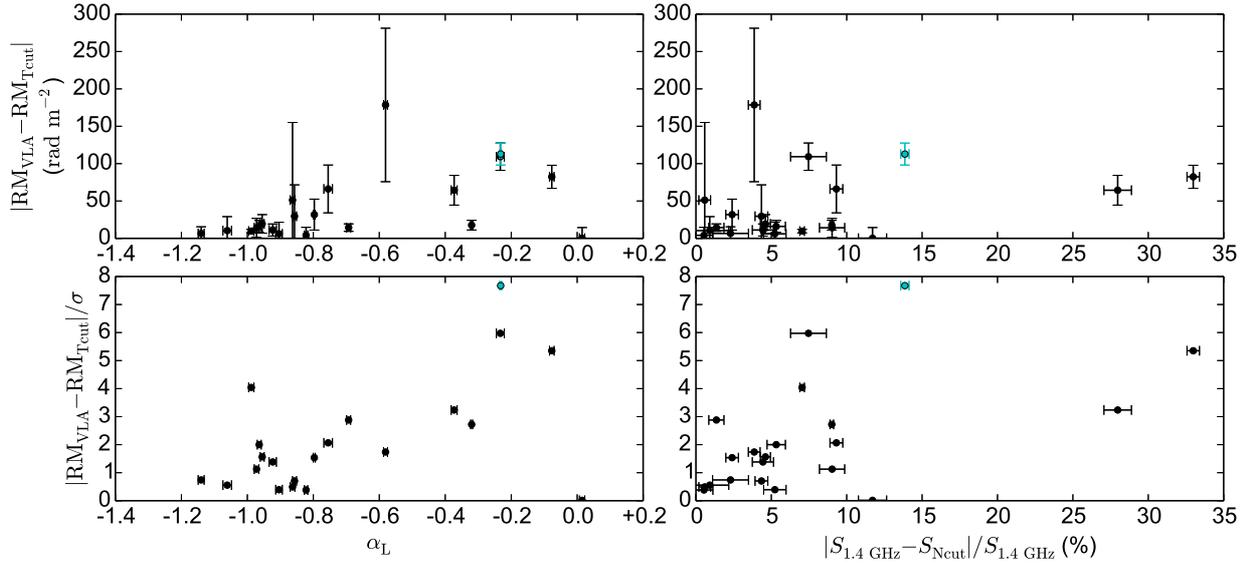}
\caption{Plots of measures of RM time variabilities against that of Stokes \textit{I} time variabilities of our 21 polarised sources. The cyan points represent J111857$+$123442, for which we used \citetalias{taylor09} cutout Stokes \textit{I} value instead of that from NVSS cutout image (see Section~\ref{sec:ivar}).\label{fig:rm_i_var}}
\end{figure*}

\subsection{Discrepancies in Total Intensities \label{sec:ivar}} 
Before comparing the flux densities between our 2014 observations and that of the NVSS in the 1990s, we note the different absolute flux density scales applied. The NVSS used 3C~295 as the flux calibrator adopting the \cite{baars77} scale, while we used 3C~286 and 3C~138 following the \cite{perley13a} scale. It has been suggested that the systematic differences between these two scales at $1.4\,{\rm GHz}$ is less than 4 per cent \citep{perley13a}, though they further noted flux density variations of $\lesssim 5$ per cent for 3C~138 over $\approx 10\,{\rm yr}$ at $1.4\,{\rm GHz}$. We have therefore chosen a modest cutoff of $10$ per cent in flux density variations above which we are confident that the differences are not due to errors in flux calibrations alone.

We compared our broadband $S_{1.4\,{\rm GHz}}$ with $S_{\rm Ncut}$ obtained from NVSS cutout images by introducing a parameter (see Table~\ref{table:cutout})
\begin{equation}
\frac{\Delta S}{S_{\rm 1.4\,GHz}} = \frac{S_{\rm 1.4\,GHz} - S_{\rm Ncut}}{S_{\rm 1.4\,GHz}}{\rm .}
\end{equation}
Four of our sources (namely, J094808$-$344010, J111857$+$123442, J170934$-$172853, and J224412$+$405715) have differences in total intensities of more than 10 per cent. These cannot be explained by flux calibration errors alone, and are likely linked to Stokes \textit{I} time variabilities between NVSS and our observations (over roughly 20 years). As all these four sources are listed as compact in the original NVSS catalogue (angular size $< 20^{\prime\prime}$)\footnote{The angular size of J111857$+$123442 was reported as $2^{\prime\prime} \times 1^{\prime\prime}$ in the FIRST survey \citep{becker95}.}, this is consistent with the general picture that the variable radio emission originates from the core of AGNs. We gathered flux density measurements of these sources in the literature up to $\sim 10\,{\rm GHz}$ and plotted them in Figure~\ref{fig:stokesi_all} to facilitate comparisons.

J111857$+$123442 shows the most extreme Stokes \textit{I} disparity of about 45 per cent when compared with its NVSS value. However, as seen in Figure~\ref{fig:stokesi_all} the NVSS flux density is inconsistent with others' as well as ours. We looked into the NVSS image containing this source (C1112P12), and noticed that at the centre of NVSS pointing 11195$+$12260, which is the closest pointing to J111857$+$123442, there is a patch with radius of $2^\prime$ where the intensity is exactly zero. Note that ($4^\circ \times 4^\circ$) NVSS images are weighted averages of their constituent snapshot images (one from each pointing; truncated at radius of $24^\prime$), with the weights defined to be proportional to the square of the primary beam attenuation \citep{condon98}. Since individual NVSS pointings are separated by $26^\prime$, the snapshot image of a missing pointing would result in a nearly circular area with $2^\prime$ radius of zero pixel values, just as we found above. This suggests that the NVSS pointing 11195$+$12260 could be missing, and the NVSS flux density of J111857$+$123442 could be unreliable due to the weighted average algorithm. To further strengthen this argument, we computed the expected resulting flux density due to the missing pointing. There is a total of three NVSS pointings covering J111857$+$123442 (11195$+$12260, 11180$+$12392, and 11195$+$12524), with the source situated $11\farcm9$, $14\farcm6$, and $19\farcm4$ from the respective pointing centres, where the primary beam attenuation levels are $0.665$, $0.541$, and $0.338$ respectively \citep[equation~5 of][]{condon98}. If we replace the source flux density in the first pointing by zero, we obtain a weighted average flux density of 47.9 per cent of the true value, exactly matching the 47.9 per cent by comparing the NVSS value with that from \citetalias{taylor09} image. Even though the \citetalias{taylor09} images were also formed using the published NVSS visibility data, they were created by mosaicking with the primary beam response divided out, and any missing pointings would lead to an increase in root-mean-square (rms) noise instead of erroneous flux densities. Indeed, the flux density of $2041.5 \pm 1.0\,{\rm mJy}$ from our broadband observation is much closer to the flux densities of $2324.4 \pm 5.5\,{\rm mJy}$ from \citetalias{taylor09} cutout images and the integrated flux density of $2322.0 \pm 0.4\,{\rm mJy}$ from the Faint Images of the Radio Sky at Twenty-Centimeters \citep[FIRST;][]{becker95} catalogue. However, there is still a 13.9 per cent discrepancy which could be true Stokes \textit{I} variability. In below, we compare our broadband $S_{1.4\,{\rm GHz}}$ with the flux density we obtained from \citetalias{taylor09} cutout for J111857$+$123442, instead of the $S_{\rm Ncut}$ from NVSS cutout images.

We first look into $\alpha_{\rm L}$ of the four sources with significant Stokes \textit{I} variabilities of more than 10 per cent. As expected, they all exhibit flat radio spectra ($-0.373 \leq \alpha_{\rm L} \leq +0.015$). We further compared the discrepancies in RM with Stokes \textit{I} variabilities (Figure~\ref{fig:rm_i_var}). It appears that flat spectrum sources are more likely to have significant RM discrepancies, though such differences in RM are not necessarily accompanied by Stokes \textit{I} variabilities. Note that the RM differences between our new observations and \citetalias{taylor09} could be attributed to off-axis polarisation leakage instead of true RM time variabilities (see Section~\ref{sec:polvar}).

\subsection{Discrepancies in Polarisation Properties} \label{sec:polvar}
By comparing the polarisation properties of our sample in \citetalias{taylor09} with that from our new VLA observations (Paper I), we found that two of the sources are unpolarised. Furthermore, the RM values of our 21 polarised target sources from our new VLA observations do not agree with that from \citetalias{taylor09} within measurement uncertainties. As we will show below, both are likely linked to off-axis instrumental polarisation in the NVSS observations.

\subsubsection{Unpolarised Sources \label{sec:weakpol}}
From the RM-Synthesis analysis on our new broadband (1--2\,GHz) data in Paper I, we found that two of our targets (J084600$-$261054 and J234033$+$133300) have polarisation fractions below our $6\sigma$ detection limits of 0.07 and 0.06 per cent, respectively. These polarisation levels are much lower than the respective values of $0.51 \pm 0.02$ and $0.59 \pm 0.02$ per cent reported in the \citetalias{taylor09} catalogue, as well as the $0.14 \pm 0.03$ and $0.09 \pm 0.02$ per cent in the original NVSS catalogue. Our on-axis broadband results are free of off-axis instrumental effects of the VLA, and are resilient against bandwidth depolarisation for sources with Faraday depths $\lesssim 10^4\,{\rm rad\,m}^{-2}$ (see Paper I). On the other hand, sources are in general placed significantly away from the pointing axis for surveys such as the NVSS, and therefore the off-axis polarisation leakage has to be taken care of for both the original NVSS catalogue \citep{condon98} and \citetalias{taylor09}. Corrections determined from holography scans were applied in the image plane in the original NVSS but not in \citetalias{taylor09}, although the off-axis leakage pattern has been smoothed out by the mosaicking done to produce the \citetalias{taylor09} images. Therefore, the residual leakage level in the original NVSS \citep[$\approx 0.3$ per cent;][]{condon98} is significantly lower than in the NVSS RM catalogue \citepalias[$\approx 0.5$ per cent;][]{taylor09}. However, the band-separated analysis of \citetalias{taylor09} has made them more robust against bandwidth depolarisation than the original NVSS.

Combining all the information above, we favour the interpretation that the polarisation signals detected from the two sources (J084600$-$261054 and J234033$+$133300) in the NVSS observations are dominated by off-axis polarisation leakage. In other words, we believe that these two sources are also unpolarised at the NVSS epoch, but were listed in the \citetalias{taylor09} catalogue just because the residual off-axis instrumental polarisation have made them appear polarised falsely. These two sources lie $10\farcm3$ and $12\farcm3$ away from the respective closest pointing centres of NVSS fields 08465$-$26188 and 23405$+$13453. At such angular distances, the (uncorrected) off-axis linear polarisation leakage is at $\sim 1$ per cent level \citep[see figure 14 of][]{condon98}, consistent with the measured values of these sources in the RM catalogue. However, we cannot completely rule out the possibility of changes in the observed polarisation fraction due to time variabilities \citep[e.g.][]{aller70,rudnick85,anderson19}.

\subsubsection{Rotation Measure Discrepancies \label{sec:rmvar}}
By comparing our results from new VLA data in NVSS bands with \citetalias{taylor09}, we found that the RM values derived from the two observations do not agree within their measurement uncertainties (see Section~\ref{sec:nvss_comp}). We first rule out the possibility that the observed RM discrepancies are due to PA calibration errors, as we do not see systematic trends in $\Delta {\rm RM}$ by grouping the target sources by the associated PA calibrators used in our observations. Another explanation to the RM disparity is genuine RM time variabilities, which can occur when a jet component near the AGN core is traversing along the jet and illuminating different parts of the foreground magnetised plasma near the jet at different epochs. Changes in RM of $\approx 100$--$1000\,{\rm rad\,m}^{-2}$ have been noted for observations at $\gtrsim 10\,{\rm GHz}$ within as short as a few months \citep[e.g.][]{zavala01,hovatta12}, and much lower values of $\approx 10\,{\rm rad\,m}^{-2}$ at 1.4\,GHz have been reported over $\approx 20\,{\rm yr}$ \citep[e.g.][]{anderson16}. This apparent variability for our targets will be further investigated with followup broadband polarisation observations (Ma et al.\ in prep). Finally, the discrepancies in measured RMs could also be attributed to some unaccounted systematic uncertainties in either or both of the observations, leading to underestimated uncertainties in RM. Here we propose that the disagreement in RM is mostly caused by residual off-axis polarisation leakage in the NVSS RM catalogue, as it can be thought of as adding a leakage vector\footnote{Complex polarisations behave as vectors in the \textit{QU}-plane, and can be simply added together. However, note that polarisation planes in the physical space do not add up as vectors (e.g.\ orthogonal polarisation planes cancel out each other).} to the source polarisation vector, modifying the measured PA values and thus the resulting RM. In the following, we quantify the effect of off-axis leakage on ${\rm RM}_{\rm TSS09}$ by simulations.

\begin{figure*}
\includegraphics[width=\doublecolumnwidth]{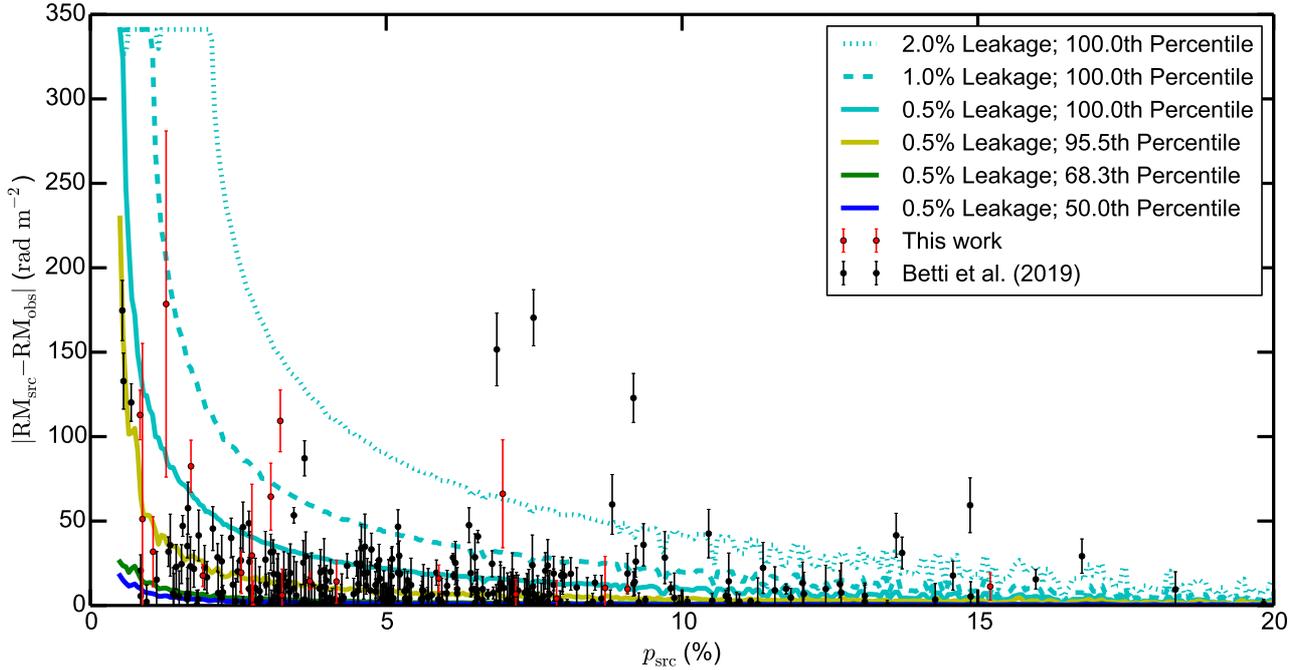}
\caption{Simulation results of the off-axis polarisation leakage effects on the measured RM values reported in the \citetalias{taylor09} NVSS RM catalogue. The $y$-axis shows the difference between true and observed RM due to the added leakage, and $x$-axis is the true polarisation percentage. With injected leakage level of 0.5 per cent of the Stokes \textit{I} flux densities, boxcar percentiles (with binning width of 0.1 per cent) from the 37,543,000 realisations were computed and shown as the colour solid lines. The 100.0th percentile (i.e.\ maximum) lines with leakage levels of 1.0 and 2.0 per cent are also shown for comparison. We also over-plot our 21 polarised target sources as red points, and the 282 Smith Cloud sources \citep{betti19} as black points. For these sources, the $y$-values are the difference in RM between \citetalias{taylor09} and new VLA results. \label{fig:sim_main}}
\end{figure*}

\begin{figure}
\includegraphics[width=\columnwidth]{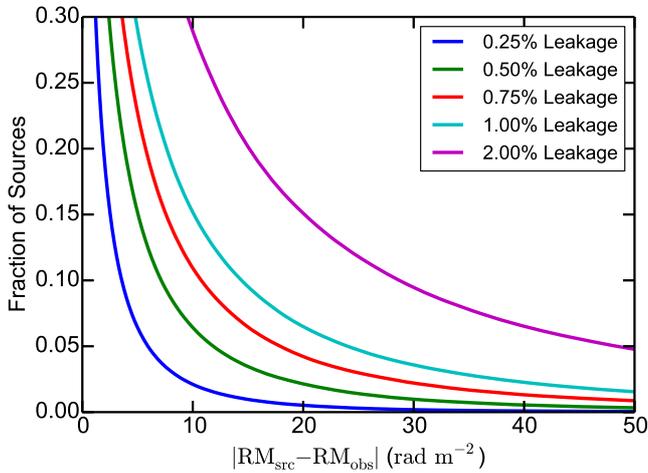}
\caption{The inverse cumulative distribution function ($1-{\rm CDF}$) of the difference in RM due to off-axis polarisation leakage from our simulation. Injected leakage levels of 0.25, 0.50, 0.75, 1.00, and 2.00 per cent are adopted and shown here. \label{fig:sim_cdf}}
\end{figure}

\begin{figure*}
\includegraphics[width=160.0pt]{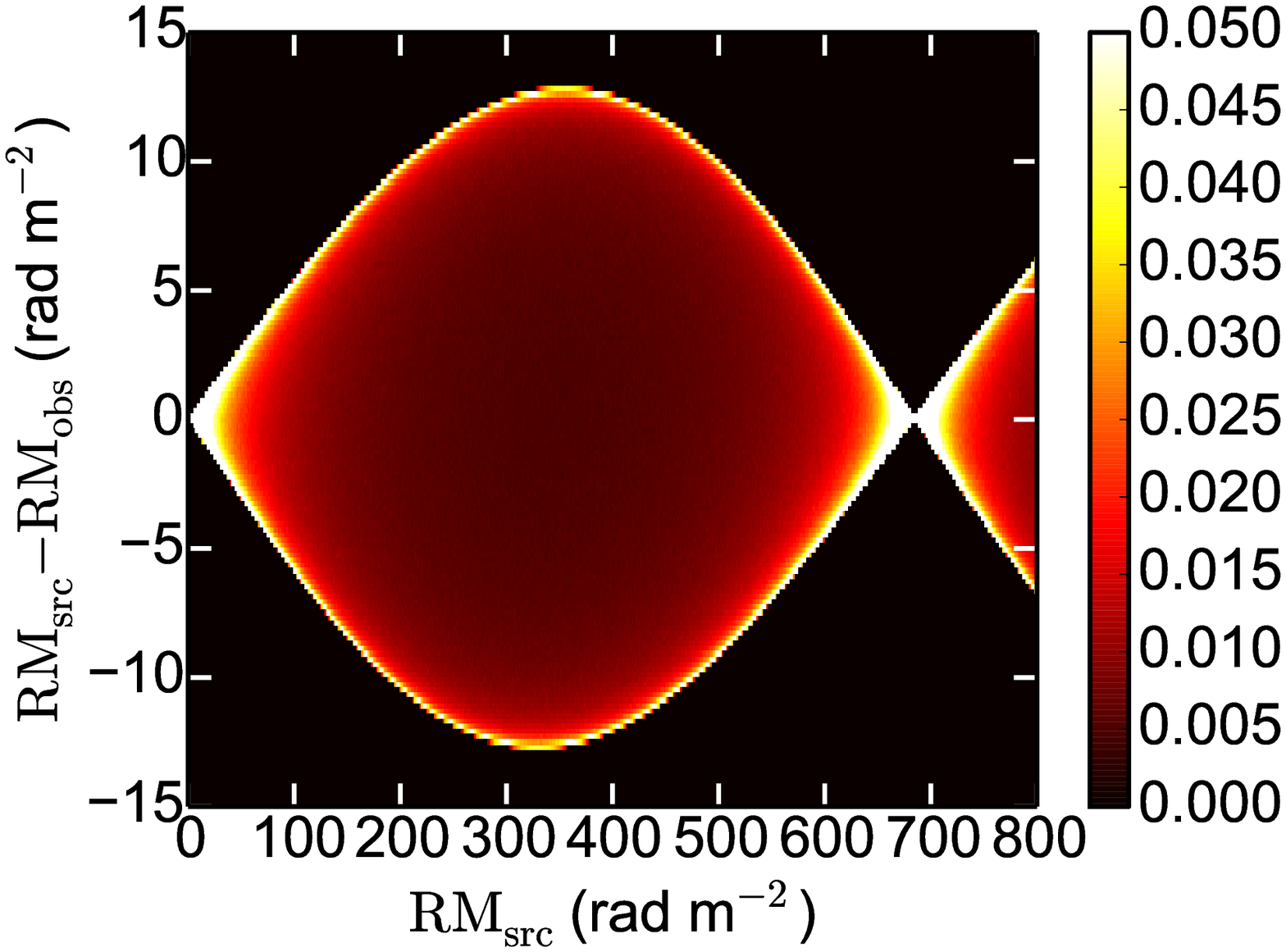}
\includegraphics[width=160.0pt]{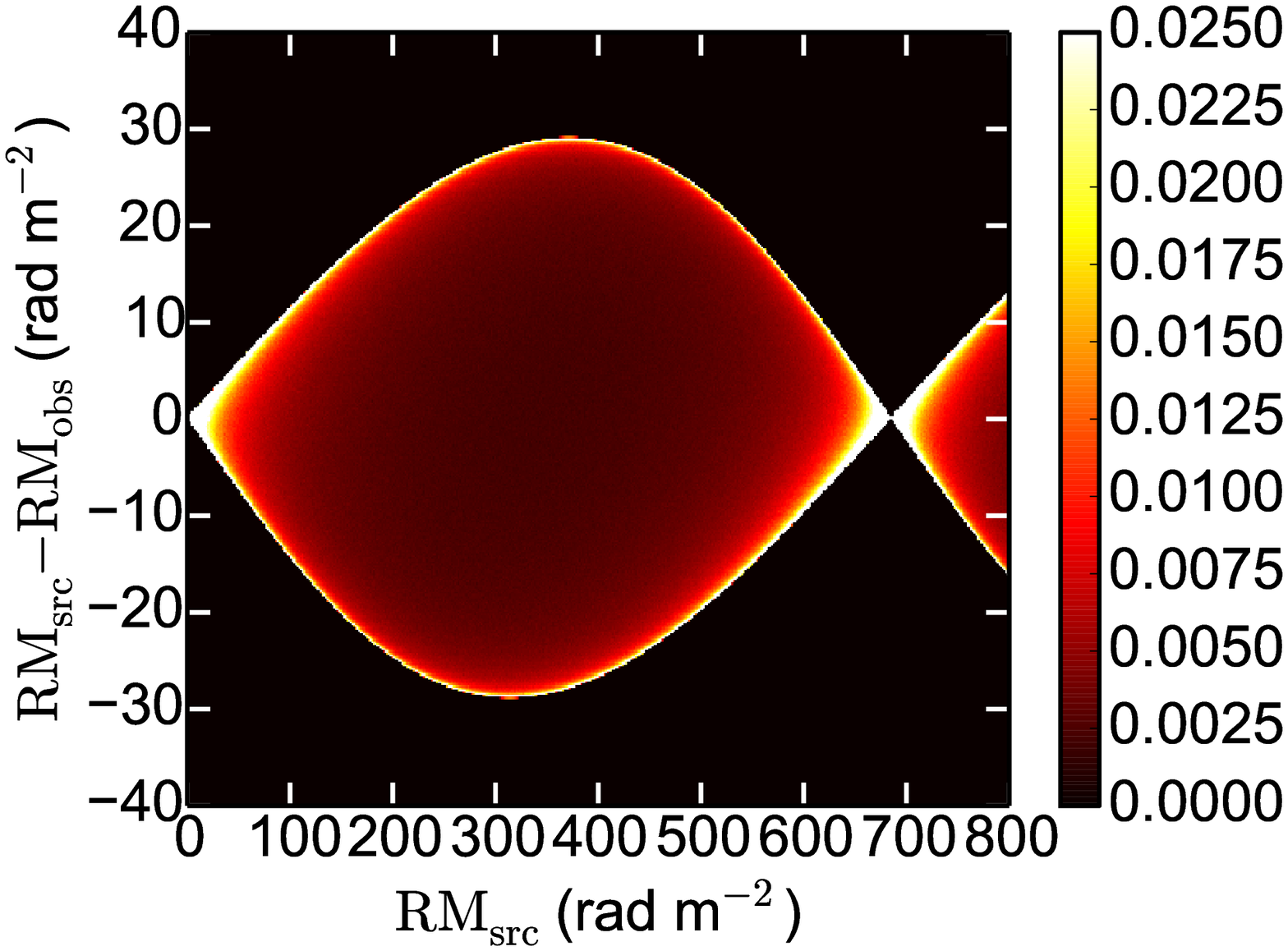}
\includegraphics[width=160.0pt]{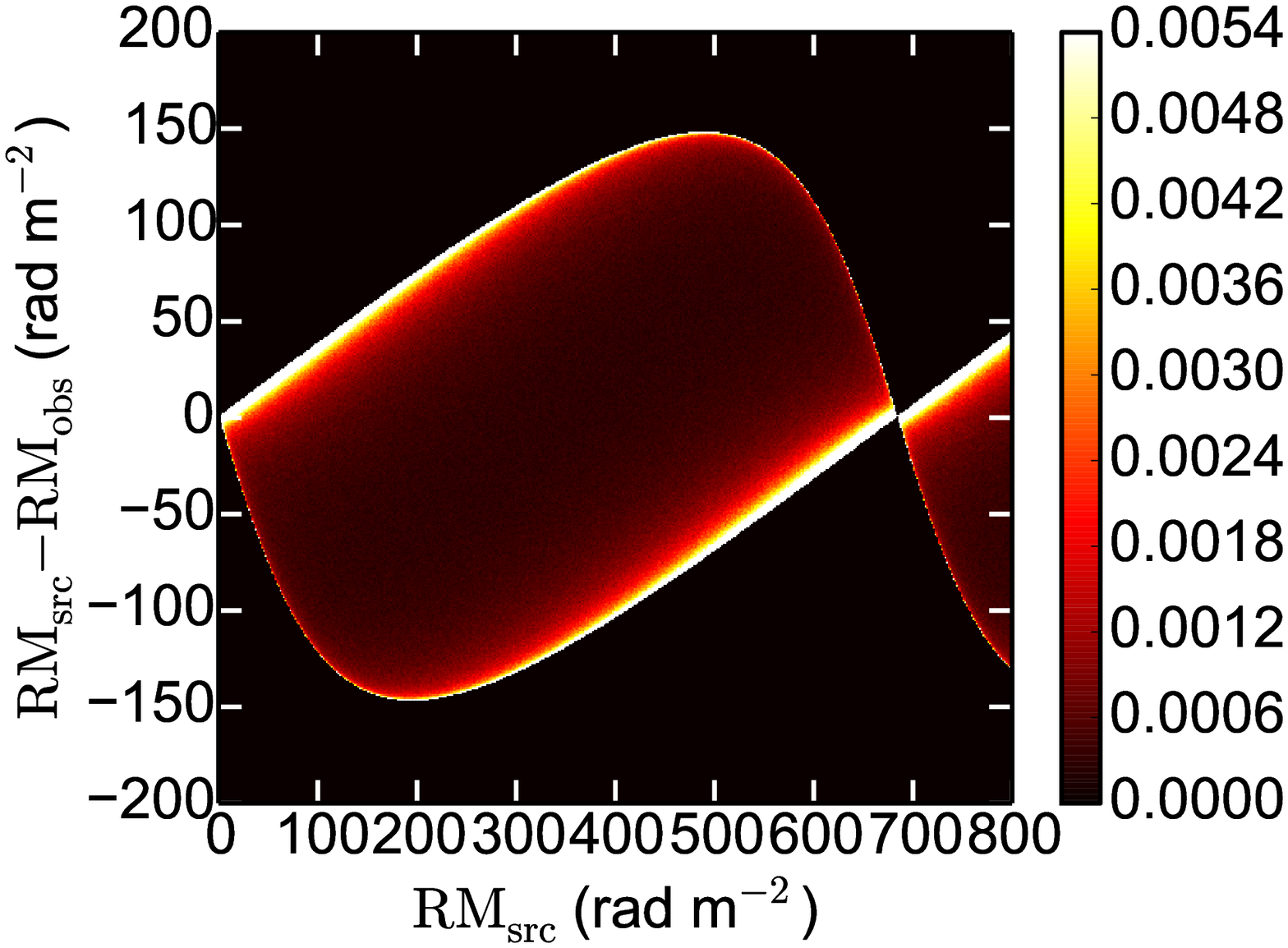}
\caption{Simulation results showing the relationship between $({\rm RM}_{\rm src} - {\rm RM}_{\rm obs})$ and ${\rm RM}_{\rm src}$ as 2D-histograms (see Section~\ref{sec:leakfrm}). The left, middle, and right panels show the cases where the artificial target sources are strongly ($p = 8.5$ per cent), intermediately ($p = 3.8$ per cent), and weakly ($p = 0.8$ per cent) polarised, respectively. Each cut along the $y$-axis at a particular ${\rm RM}_{\rm src}$ represents the artificial source with that corresponding ${\rm RM}_{\rm src}$, chosen at a $1\,{\rm rad\,m}^{-2}$ interval from $0$ to $+800\,{\rm rad\,m}^{-2}$. The same binning width of $0.2\,{\rm rad\,m}^{-2}$ along the $y$-axis has been used for all three panels, and we have normalised the histogram along each $y$-cut. Note that the $y$-axis and colour bar scales are different among the panels. \label{fig:2dhist}}
\end{figure*}

\begin{figure*}
\includegraphics[width=160.0pt]{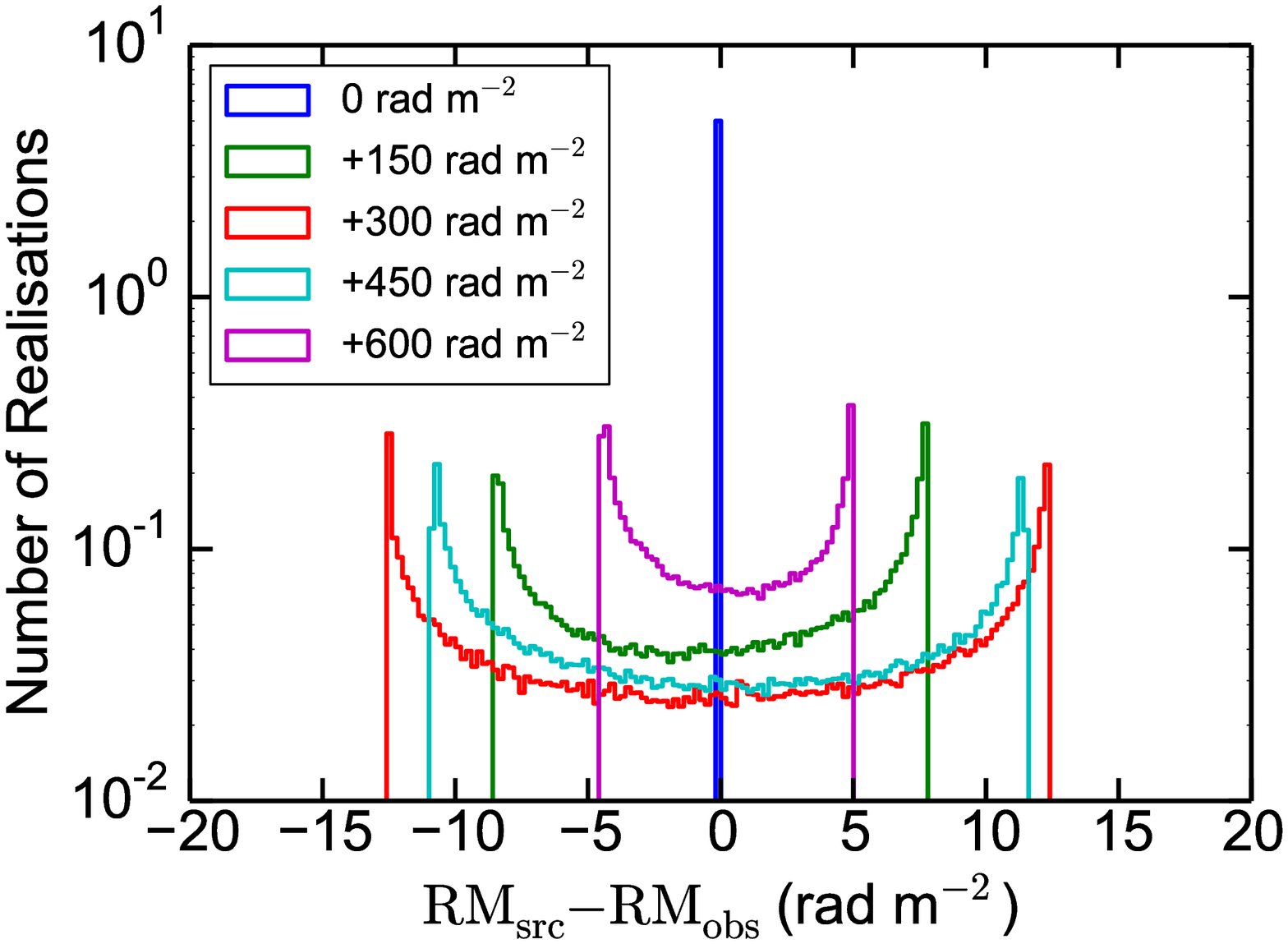}
\includegraphics[width=160.0pt]{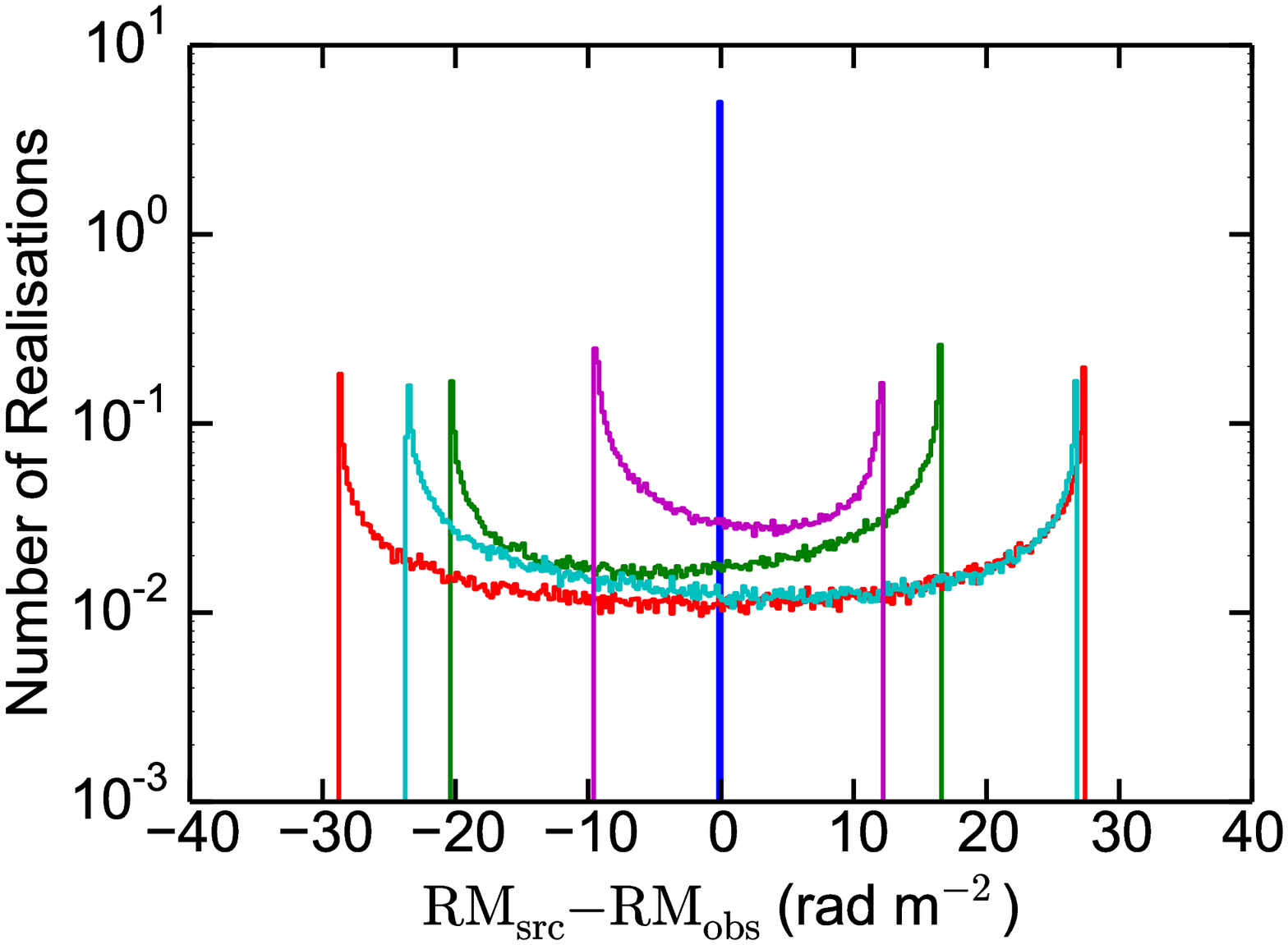}
\includegraphics[width=160.0pt]{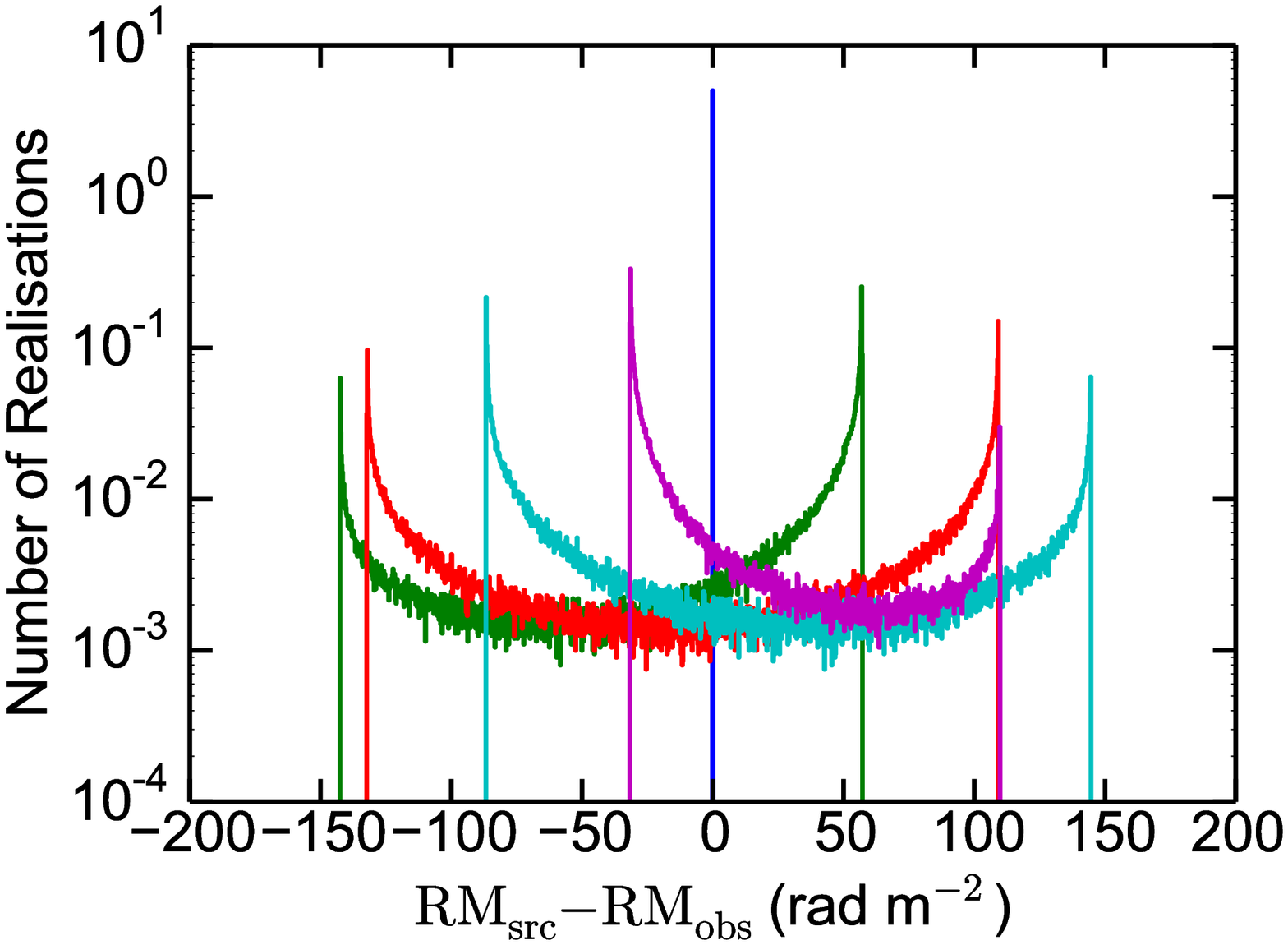}
\caption{Simulation results showing the relationship between $({\rm RM}_{\rm src} - {\rm RM}_{\rm obs})$ and ${\rm RM}_{\rm src}$ (see Section~\ref{sec:leakfrm}), plotting cuts along the $y$-axis from Figure~\ref{fig:2dhist} at ${\rm RM}_{\rm src}$ of $0$, $+150$, $+300$, $+450$, and $+600\,{\rm rad\,m}^{-2}$ as the blue, green, red, cyan, and magneta lines, respectively. Note that the $x$- and $y$-axis scales are different among the panels, and the $y$-axis is in logarithmic scale. \label{fig:1dhist}}
\end{figure*}

\section{Simulating the Effects of Off-axis Instrumental Polarisation} \label{sec:sim}
\subsection{Simulation Setup \label{sec:simsetup}}
We took the reported radio properties of the 37,543 sources in the NVSS RM catalogue to construct the source true polarisation vectors at the two NVSS bands by
\begin{equation}
\mathbf{P}_{j, {\rm src}} = Q_{j, {\rm src}} + iU_{j, {\rm src}} = {\rm PI}_{\rm src} \cdot e^{2i({\rm PA}_0 + {\rm RM}_{\rm src} \cdot \lambda_j^2)}{\rm ,}
\end{equation}
where the subscript $j$ denotes the two NVSS IFs, and $\lambda_{\rm j}$ is the centre (in frequency space) wavelength of the two IFs with $\lambda_1 = 0.2196\,{\rm m}$ and $\lambda_2 = 0.2089\,{\rm m}$. We adopted the polarised intensity listed in the \citetalias{taylor09} catalogue as ${\rm PI}_{\rm src}$ and the RM from \citetalias{taylor09} as ${\rm RM}_{\rm src}$. Although we have set up the simulation by taking the listed \citetalias{taylor09} values as the source true values while in reality they should be taken as the observed values with polarisation leakage added in, we argue that our simulation results would still be representative of the general statistics of the \citetalias{taylor09} catalogue given its large sample size. A full treatment taking the listed \citetalias{taylor09} values as the observed values from our simulations will be presented in the future Paper III. It is assumed that all the sources have flat spectral indices ($\alpha_{\rm L} = 0$) and are Faraday simple. Since ${\rm PA}_0$ were not reported in the \citetalias{taylor09} catalogue, we constructed 1,000 realisations for each source, each with a randomly picked ${\rm PA}_0$ value within $[-\pi/2$, $+\pi/2]$ from a uniform distribution. This results in a total of 37,543,000 input realisations.

Then, we added leakage vectors ($\mathbf{P}_{\rm leak}$) to the true polarisation vectors to obtain the observed polarisation vectors:
\begin{align}
\mathbf{P}_{j, {\rm obs}} &= \mathbf{P}_{j, {\rm src}} + \mathbf{P}_{\rm leak}{\rm ,}\\
\mathbf{P}_{\rm leak} &= (S_{\rm NVSS} \times 0.5\,\%) \cdot e^{2i{\rm PA}_{\rm leak}}{\rm .}
\end{align}
The leakage vector has an amplitude fixed at 0.5 per cent of the NVSS Stokes \textit{I} values, and ${\rm PA}_{\rm leak}$ again randomly picked within $[-\pi/2$, $+\pi/2]$ from uniform distribution for each realisation, identical at IF1 and IF2. In other words, it is assumed that the leakage amplitude does not depend on the source position within the telescope primary beam with respect to the pointing centre, which is justifiable as the mosaicking done to produce the \citetalias{taylor09} images is expected to smooth out the leakage amplitude within the primary beam compared to the original leakage pattern of \cite{cotton94}. Furthermore, it is assumed that the instrumental polarisation (both amplitude and PA) is not frequency dependent, although the VLA off-axis leakage pattern in the NVSS configurations are in reality weakly dependent of both direction and frequency \citep{condon98,cotton94}. The 0.5 per cent leakage level we adopted is motivated by the \citetalias{taylor09} reported polarisation fractions of the two unpolarised sources that we identified (J084600$-$261054 and J234033$+$133300; see Section~\ref{sec:weakpol}). As a reference, if the NVSS off-axis polarisation leakage is left completely uncorrected, sources will experience leakage levels of about 0.3, 0.6, 1.6, and 2.4 per cent at distances of $5^\prime$, $10^\prime$, $15^\prime$, and $20^\prime$ away from the pointing centre respectively \citep[see table~1 of][]{cotton94}, and \citetalias{taylor09} sources have an average offset from the closest pointing centre by about $9\farcm5$. We therefore argue that an input leakage level of 0.5 per cent is a reasonable value to adopt, though note again that the residual leakage pattern in \citetalias{taylor09} is expected to be smoothed out from mosaicking. 

With the resulting polarisation vectors after adding in polarisation leakages, we computed the observed RM values (${\rm RM}_{\rm obs}$) for each of the 37,543,000 realisations by Equations~\ref{eq:2band_rm} and \ref{eq:rm}. The $n\pi$-ambiguity is resolved by choosing the closest possible RM value to ${\rm RM}_{\rm src}$.

\subsection{Unaccounted RM Uncertainties Due to Off-axis Leakage \label{sec:delrm}}

To quantify the effect of off-axis polarisation leakage on the measured RM, we compared $|{\rm RM}_{\rm src} - {\rm RM}_{\rm obs}|$ against the true polarisation fraction ($p_{\rm src} = {\rm PI}_{\rm src}/S_{\rm NVSS}$) for the 37,543,000 simulation realisations. The RM differences reflect how much the injected leakage vectors alter the source true RM values. To clearly see the underlying statistics, instead of plotting each of the realisations, we performed boxcar binning of these data points with a binning width of 0.1 per cent in $p_{\rm src}$, with the results at the 50.0th, 68.3th, 95.5th, and 100.0th percentiles plotted as the colour solid lines in Figure~\ref{fig:sim_main}. We also tested different binning widths (0.05, 0.2, and 0.5 per cent) to ensure that they all show consistent results. As expected, sources with lower fractional polarisation are more susceptible to changes in RM due to the injected polarisation leakage. At true polarisation level of 1 per cent, the RM difference at 50th percentile (i.e.\ median) is $7.3\,{\rm rad\,m}^{-2}$, while that at 68.3th percentile (corresponding to $1\sigma$ significance) is $13.5\,{\rm rad\,m}^{-2}$. These numbers are comparable to the median RM uncertainties of $10.8\,{\rm rad\,m}^{-2}$ reported in the \citetalias{taylor09} catalogue. This means the leakage effect introduces a significant extra RM uncertainty which has not been accounted for in the \citetalias{taylor09} RM catalogue, and therefore we suggest that care has to be taken when using the reported RM values of individual sources with $p \lesssim 1$ per cent.

\begin{figure}
\includegraphics[width=\columnwidth]{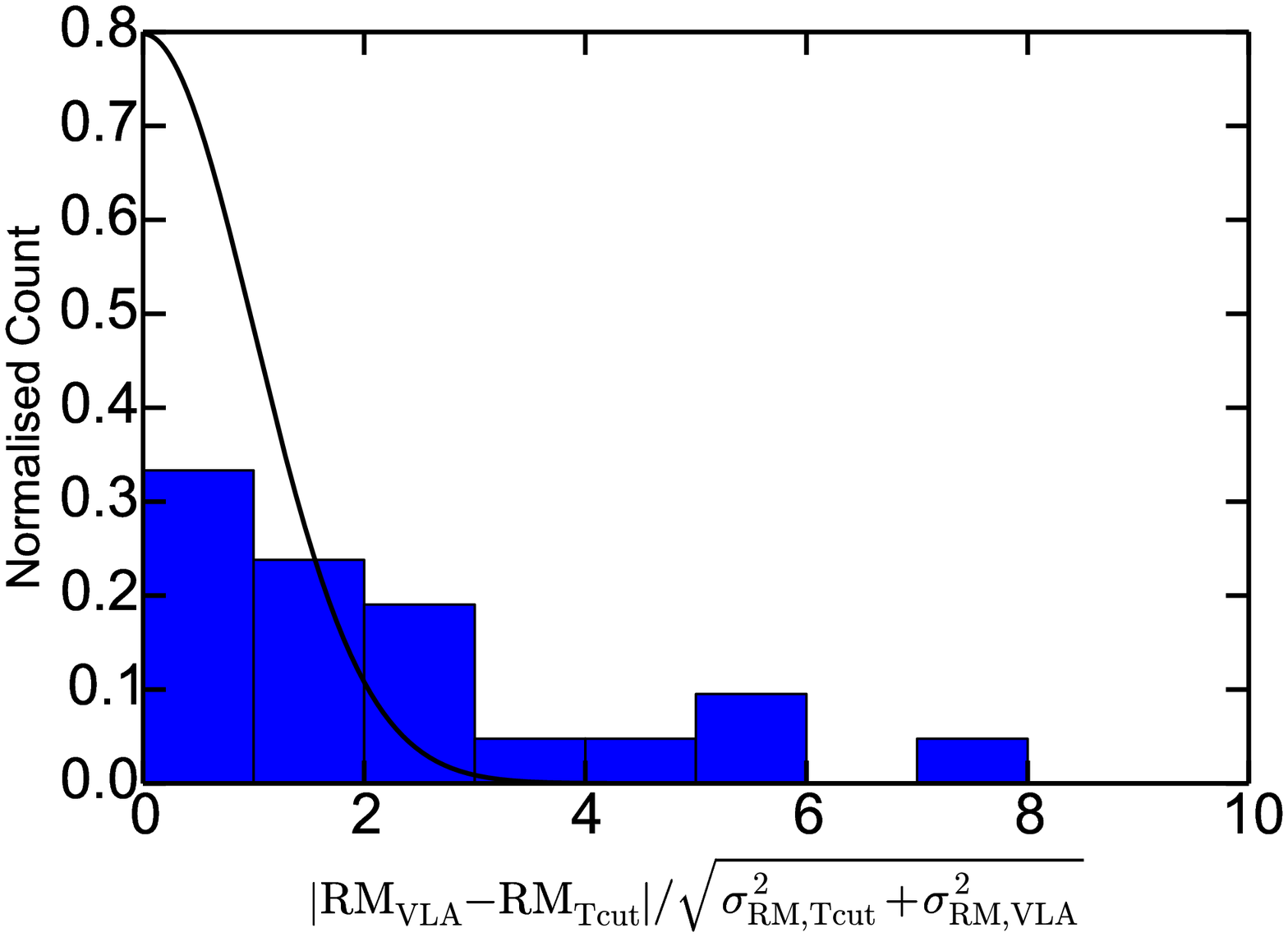}
\includegraphics[width=\columnwidth]{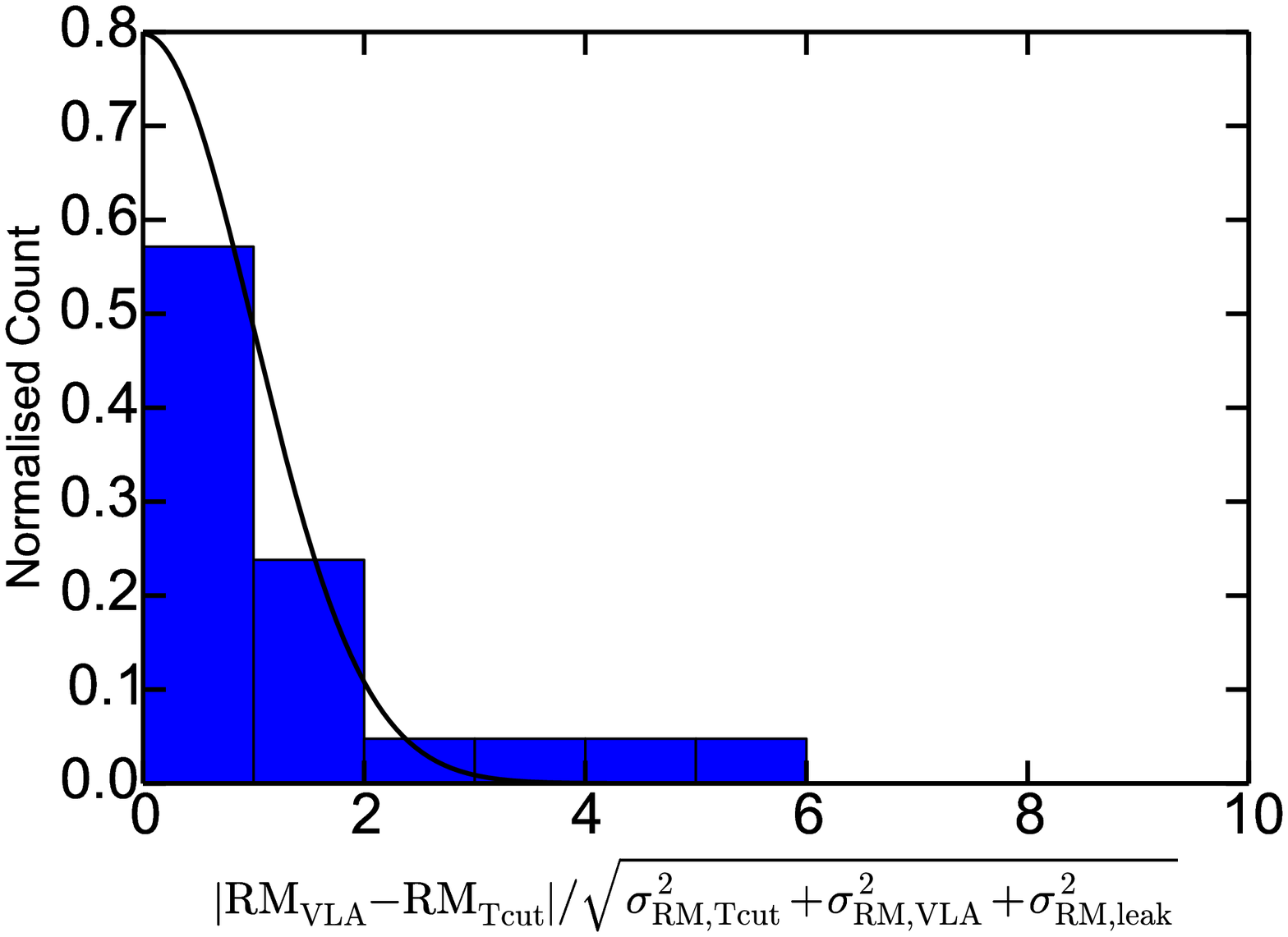}
\caption{Histograms of RM differences between our new VLA observations and that from \citetalias{taylor09} cutout images, in units of RM uncertainties. The upper plot is in units of $\sqrt{\sigma_{\rm RM, Tcut}^2 + \sigma_{\rm RM, VLA}^2}$ (i.e.\ without taking off-axis leakage into account), and the lower plot is in units of $\sqrt{\sigma_{\rm RM, Tcut}^2 + \sigma_{\rm RM, leak}^2 + \sigma_{\rm RM, VLA}^2}$ (i.e. with leakage taken into account). The black curves in both plots show a folded normal distribution with $\mu = 0$ and $\sigma = 1$. \label{fig:rmhist}}
\end{figure}

\begin{figure*}
\includegraphics[width=\doublecolumnwidth]{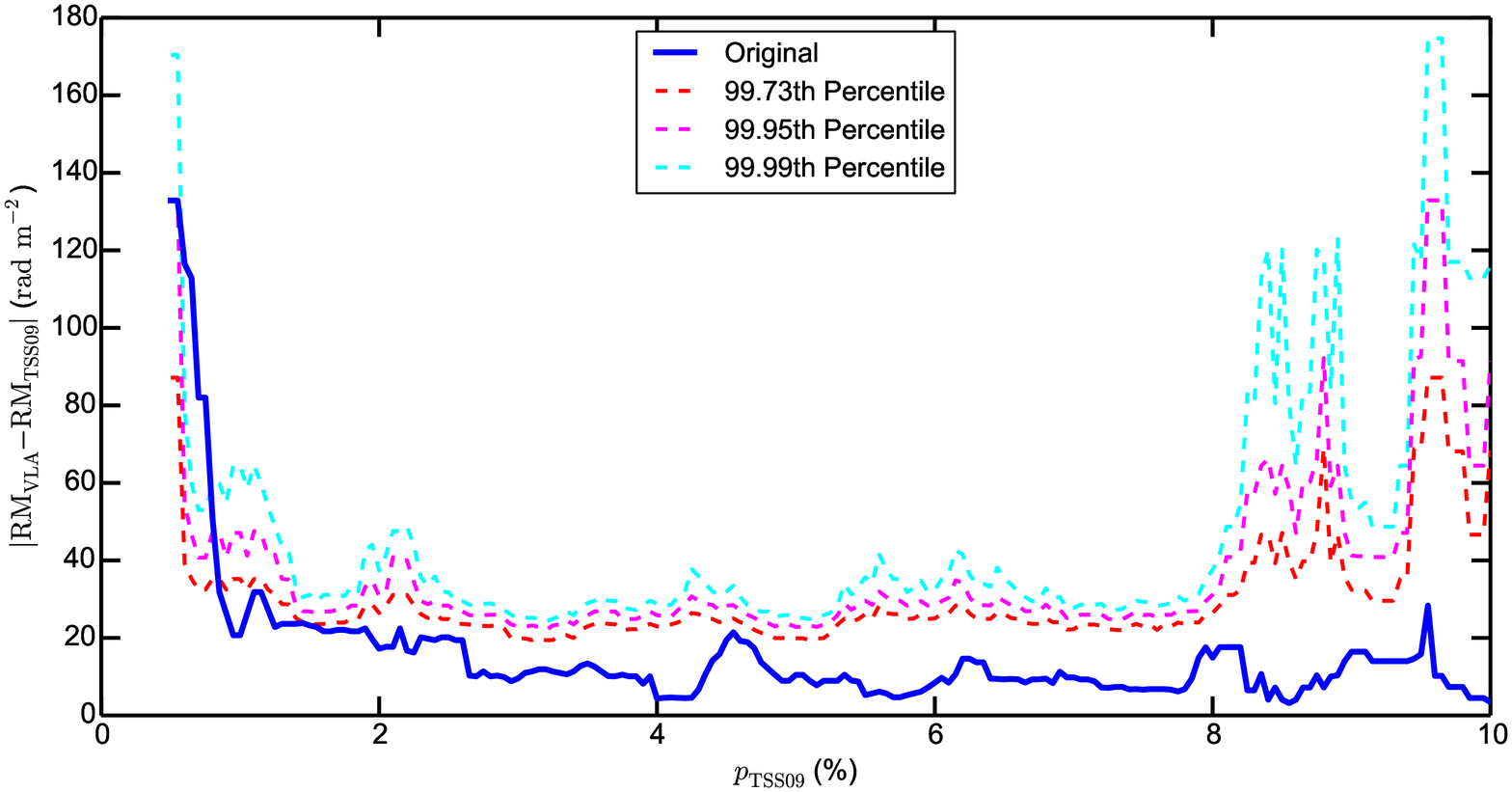}
\caption{Difference in RM between \citetalias{taylor09} and new VLA observations. The boxcar median of the 303 sources \citep[21 from this work, plus 282 from][]{betti19} is shown by the blue solid line. The colour dashed lines are the 99.73th, 99.95th, and 99.99th percentile lines respectively from $10^7$ shuffles of the $y$- with respective to the $x$-values (see Section~\ref{sec:simobs}). A boxcar binning width of 0.4 per cent is used here. \label{fig:sim_shuffle}}
\end{figure*}

The above simulation is also repeated by using different leakage levels (0.25, 0.75, 1.00, and 2.00 per cent). The 100.0th boxcar percentile lines for 1.00 and 2.00 per cent, which shows the maximum possible deviation of the observed RM from the true value at the respective leakage levels, are also shown in Figure~\ref{fig:sim_main}. The flattening of these two curves at $p_{\rm src} \lesssim 2$ per cent are in the regime where the injected leakage overpowers the true polarisation signal, and should be ignored. We also show the inverse cumulative distribution function of $|{\rm RM}_{\rm src} - {\rm RM}_{\rm obs}|$ in Figure~\ref{fig:sim_cdf}, from which we found that 2, 6, 11, 15, and 29 per cent of the NVSS RM sources have $|{\rm RM}_{\rm src} - {\rm RM}_{\rm obs}| \geq 10\,{\rm rad\,m}^{-2}$ at injected leakage levels of 0.25, 0.50, 0.75, 1.00, and 2.00 per cent respectively.

We further estimate how much \citetalias{taylor09} have underestimated their RM uncertainties for not having the off-axis leakage effect taken into account. To achieve this, we determined the rms value in $|{\rm RM}_{\rm src} - {\rm RM}_{\rm obs}|$ from the 1,000 simulation realisation for each of the 37,543 sources individually (denoted as $\sigma_{\rm RM,leak}$). This is then added in quadrature to the listed \citetalias{taylor09} RM uncertainties of the corresponding source ($\sigma_{\rm RM,TSS09}$) to yield the new RM uncertainties including the effect of off-axis leakage for each of the 37,543 sources ($\sigma_{\rm RM,new} = \sqrt{\sigma_{\rm RM,TSS09}^2 + \sigma_{\rm RM,leak}^2}$). We calculated $\sigma_{\rm RM,new}/\sigma_{\rm RM,TSS09}$ at our default injected leakage level of 0.50 per cent, and obtained a median value of 1.09 out of the entire sample of 37,543 \citetalias{taylor09} sources. In other words, the \citetalias{taylor09} RM uncertainties should be increased by an average of nine per cent to incorporate the effect of off-axis polarisation leakage. Choosing different leakage levels of 0.25, 0.75, 1.00, and 2.00 per cent would yield median $\sigma_{\rm RM,new}/\sigma_{\rm RM,TSS09}$ of 1.03, 1.14, 1.19, and 1.39 instead.

Finally, we stress here that one should be cautious when applying the above results to individual \citetalias{taylor09} sources, since we find that the RM uncertainty from this off-axis polarisation leakage is also dependent on the actual RM of the source (Section~\ref{sec:leakfrm}). The statistics reported above are the average values out of the entire RM catalogue. Another point to note is that the distribution of $({\rm RM}_{\rm src} - {\rm RM}_{\rm obs})$ is found to be asymmetric, and is highly non-Gaussian (see Section~\ref{sec:leakfrm}).

\subsection{Leakage RM Uncertainties Dependence on $\mathbf{RM}_\mathbf{src}$ \label{sec:leakfrm}}

To investigate the relationship between the RM uncertainties due to off-axis leakage and the source true RM values, we repeated the simulation outlined in Section~\ref{sec:simsetup} but with manually selected ${\rm RM}_{\rm src}$ values instead. In particular, we adopted the source properties of one of our targets, J091145$-$301305, as per \citetalias{taylor09} catalogue ($S_{\rm NVSS} = 247.1\,{\rm mJy}$ and ${\rm PI} = 21.1\,{\rm mJy}$) \emph{except} for the RM value. Instead, we have manually put in ${\rm RM}_{\rm src}$ values of $0$ to $+800\,{\rm rad\,m}^{-2}$ at $1\,{\rm rad\,m}^{-2}$ interval, resulting in 801 artificial sources. We chose J091145$-$301305 because it is strongly polarised ($p = 8.5$ per cent). We generated $10^5$ simulation realisations for each artificial source (each with randomised ${\rm PA}_0$ and ${\rm PA}_{\rm leak}$ as before; see Section~\ref{sec:simsetup}), and plotted the distribution of the resulting $({\rm RM}_{\rm src} - {\rm RM}_{\rm obs})$ individually for each artificial source as a 2D-histogram in the left panel of Figure~\ref{fig:2dhist}. We further repeated the above by using the \citetalias{taylor09} source properties of two of our other targets -- J093349$-$302700 ($S_{\rm NVSS} = 272.9\,{\rm mJy}$ and ${\rm PI} = 10.3\,{\rm mJy}$) as an intermediately polarised example ($p = 3.8$ per cent) and J235728$+$230226 ($S_{\rm NVSS} = 624.6\,{\rm mJy}$ and ${\rm PI} = 5.0\,{\rm mJy}$) as a weakly polarised example ($p = 0.8$ per cent). Both of these cases are also shown in Figure~\ref{fig:2dhist} as the middle and right panels, respectively.

From these simulations, we noted several interesting properties of the distribution of $({\rm RM}_{\rm src} - {\rm RM}_{\rm obs})$ due to the injected off-axis polarisation leakage. Firstly, the distribution is a strong function of ${\rm RM}_{\rm src}$, with the spread in $({\rm RM}_{\rm src} - {\rm RM}_{\rm obs})$ being the widest at ${\rm RM}_{\rm src} \sim 300\,{\rm rad\,m}^{-2}$, while it is identical to zero for the case of ${\rm RM}_{\rm src} = 0$ and $\pm 682.3\,{\rm rad\,m}^{-2}$ (i.e., the injected leakage would not alter the RM value at all). These reported numbers, however, are only valid under the assumption that ${\rm PA}_{\rm leak}$ are identical in the two IFs. Otherwise (i.e., if the off-axis leakage has a non-zero RM), the distributions shown in Figure~\ref{fig:2dhist} will be shifted horizontally by the leakage RM. Secondly, the distribution at any given ${\rm RM}_{\rm src}$ (except for the identical zero cases mentioned above) is highly non-Gaussian, shaped as a double horn. This can be seen more clearly in Figure~\ref{fig:1dhist} where we show cuts of the 2D-histograms at ${\rm RM}_{\rm src}$ of $0$, $+150$, $+300$, $+450$, and $+600\,{\rm rad\,m}^{-2}$. An interesting implication of this is that, the listed ${\rm RM}_{\rm TSS09}$ values are in general \emph{not} the most likely true RM of the sources, even for the highly polarised example with $p = 8.5$ per cent. Nonetheless, we find that the mean of the $({\rm RM}_{\rm src} - {\rm RM}_{\rm obs})$ distribution at any given ${\rm RM}_{\rm src}$ is almost identical to $0\,{\rm rad\,m}^{-2}$ for all cases. Finally, it can be seen that the distributions are asymmetric, with this property being more pronounced for sources with lower fractional polarisation. We will briefly explore the underlying causes of these properties in Appendix~\ref{sec:paleak_sim}.

\subsection{Comparing Simulation with New Observations \label{sec:simobs}}
We return to comparing the RM from our new observations (${\rm RM}_{\rm VLA}$) with that from \citetalias{taylor09} cutout images (${\rm RM}_{\rm Tcut}$) after taking this leakage effect into account. For each of our polarised target sources, we computed the RM uncertainty due to this off-axis leakage ($\sigma_{\rm RM, leak}$) by repeating the simulation (Section~\ref{sec:simsetup}) but only considering our targets one at a time, and with a much higher number of simulation realisation of $10^5$ per source. The rms of the resulting $|{\rm RM}_{\rm src} - {\rm RM}_{\rm obs}|$ values is taken as the $\sigma_{\rm RM, leak}$ of that source. We reiterate here that the distribution of $({\rm RM}_{\rm src} - {\rm RM}_{\rm obs})$ is highly non-Gaussian (see Section~\ref{sec:leakfrm}).

The RM discrepancies between the two observations are plotted in the form of histograms in Figure~\ref{fig:rmhist}, in units of $\sqrt{\sigma_{\rm RM, Tcut}^2 + \sigma_{\rm RM, VLA}^2}$ (i.e.\ without taking leakage into account), as well as $\sqrt{\sigma_{\rm RM, Tcut}^2 + \sigma_{\rm RM, VLA}^2 + \sigma_{\rm RM, leak}^2}$ (i.e.\ with leakage taken into account), with $\sigma_{\rm RM, Tcut}$ and $\sigma_{\rm RM, VLA}$ being the RM uncertainties in \citetalias{taylor09} cutout and our VLA observations, respectively. If the difference in RM is only due to random Gaussian noise, the histogram should follow a folded normal distribution with $\mu = 0$ and $\sigma = 1$, which is shown as the black curves in both plots. Indeed, the histogram follows the expected distribution much more closely after taking into consideration the effect of the off-axis leakage. This means the RM discrepancies between our new VLA observations with that from \citetalias{taylor09} can be largely explained by the off-axis instrumental polarisation leakage at about 0.5 per cent level. The three sources still with significant RM discrepancies even after considering the off-axis leakage are J084701$-$233701 (at $5.0\sigma$), J111857$+$123442 (at $4.8\sigma$), and J170934$-$172853 (at $3.5\sigma$). This could be due to genuine RM time variabilities, and will be investigated in a forthcoming paper (Ma et al.\ in prep).

\subsection[]{Additional Evidence from the \cite{betti19} Data}

To further confirm that the RM discrepancies are due to off-axis polarisation leakage, we supplemented our 21 sources with the 282 Smith Cloud sources from \cite{betti19} that are cross-matched with the \citetalias{taylor09} catalogue. Their sources were also placed on-axis in their broadband VLA observations in L-band and are thus unaffected by off-axis polarisation leakage. However, there are significant differences between their data and that of \citetalias{taylor09}, namely (1) they used broadband 1--2\,GHz data while \citetalias{taylor09} used narrowband observations, (2) their observations were conducted in A array configuration instead of D array as did NVSS, and (3) their RM values were taken as the peak from RM-Synthesis analysis, while \citetalias{taylor09} performed two-point $\lambda^2$ fit to PA. Such differences in $\lambda^2$ coverage, \textit{uv}-coverage, and analysis method can lead to disagreements between their RM values and the corresponding \citetalias{taylor09} RM values, which could be mistaken as caused by the off-axis polarisation leakage of \citetalias{taylor09} and/or RM time variabilities. A detailed, quantitative comparison between \cite{betti19} and \citetalias{taylor09} is therefore outside the scope of this work.

The combined 303 sources are plotted in Figure~\ref{fig:sim_main}. We computed the boxcar median line for these 303 observed sources, as shown in Figure~\ref{fig:sim_shuffle} (denoted as ``original''). A boxcar binning width of 0.4 per cent is used here instead of the 0.1 per cent we used in Figure~\ref{fig:sim_main} because of the lack of observed sources in some $p$ bins. We do not quantitatively compare this observed median line with the simulation results (Figure~\ref{fig:sim_main}) directly, because the selection biases of the two observations preclude meaningful conclusions to be drawn. In particular, our target sources were selected to have $|{\rm RM}_{\rm TSS09}| \gtrsim 300\,{\rm rad\,m}^{-2}$, while the Betti et al.\ sources are in close proximity to the Smith Cloud and therefore have a different RM distribution than that of the entire sample of \citetalias{taylor09}. As the distribution of RM uncertainties due to off-axis leakage is a function of the source RM (Section~\ref{sec:leakfrm}), it would not be surprising if the median line here does not closely match the simulation results presented above in Section~\ref{sec:delrm}.

We focus here on the qualitative trend of the observed median line, from which we find a peak at the lower end of $p$ ($\lesssim 1$ per cent). This is similar to what we see from our simulation, and is an important indicator that the discrepancies in RM are mainly due to off-axis polarisation leakage. However, this peak may also simply be a result of small number statistics -- sources with larger $|{\rm RM}_{\rm VLA} - {\rm RM}_{\rm TSS09}|$ might just have low $p$ coincidentally. In such case, $|{\rm RM}_{\rm VLA} - {\rm RM}_{\rm TSS09}|$ would actually have no correlation with $p$. We test this hypothesis with the bootstrapping method. With the 303 observed sources, we shuffle the $|{\rm RM}_{\rm VLA} - {\rm RM}_{\rm TSS09}|$ values with respect to $p$, and then construct a new boxcar median line (also at binning width of 0.4 per cent). This shuffling process is repeated for $10^7$ times, yielding $10^7$ median lines. As the final step, for each $p$ value we evaluated the 99.73th, 99.95th, and 99.99th percentiles of $|{\rm RM}_{\rm VLA} - {\rm RM}_{\rm TSS09}|$ of the $10^7$ median lines.

These percentile lines are plotted in Figure~\ref{fig:sim_shuffle} together with the original observed median line. The percentile lines peak at several $p$ values -- at 0.5 per cent and between 8.0 and 10.0 per cent. These peaks are due to the scarcity of data points there, as can be seen in Figure~\ref{fig:sim_main}. The original line is mostly featureless and lie well below the 99.73th percentile line for $p \gtrsim 1$ per cent, while for $p \lesssim 1$ per cent it peaks to a level close to the 99.95th percentile line. This suggests that the trend of higher $|{\rm RM}_{\rm VLA} - {\rm RM}_{\rm TSS09}|$ at lower $p$ is statistically robust (at about 99.95 per cent confidence level), and that residual off-axis polarisation leakage is indeed introducing extra RM uncertainties to \citetalias{taylor09}.

\subsection[]{Comparison with \cite{stil11}}

When \cite{stil11} compared the RM structure function of the Galactic poles they derived from \citetalias{taylor09} results with that from \cite{mao10}, they suggested that the RM uncertainties in \citetalias{taylor09} catalogue could be underestimated by a factor 1.22. This would explain the discrepancies between the two studies, though the authors did not explore the cause of such underestimated uncertainties. As we have shown above, off-axis polarisation leakage can introduce extra RM uncertainties to \citetalias{taylor09} results, and can potentially explain the factor 1.22 that they suggested. 

To test this, we compiled the list of sources taken as the North Galactic Pole (NGP; 1,019 sources) and South Galactic Pole (SGP; 752 sources) samples by \cite{stil11}, and investigated the $\sigma_{\rm RM,new}/\sigma_{\rm RM,TSS09}$ ratio of these sources. This ratio quantifies by what factor we should increase the \citetalias{taylor09} RM uncertainties in order to incorporate the effects of off-axis polarisation leakage, and were obtained in Section~\ref{sec:delrm}. Using our default leakage level of 0.5 per cent, we find that the median values of $\sigma_{\rm RM,new}/\sigma_{\rm RM,TSS09}$ for the NGP and SGP sources are both 1.02, much lower than the factor of 1.22 suggested by \cite{stil11}. This suggests that there can be other sources of \citetalias{taylor09} RM uncertainties that have not been accounted for yet.

\section{Conclusion} \label{sec:conclusion}
From our new broadband spectro-polarimetric observations of 23 NVSS RM sources with the VLA, we identified two unpolarised sources that are listed as $\approx 0.5$ per cent polarised in the \citetalias{taylor09} catalogue. Moreover, we found significant discrepancies in RM for the remaining 21 sources by carefully comparing our new data with that from \citetalias{taylor09} using the same analysis methods and in almost identical frequency ranges. We attributed both of these effects to the residual off-axis instrumental polarisation in the \citetalias{taylor09} catalogue. We quantified its effects on the measured RM using simulations, and found that it is more significant for sources with lower fractional polarisation, leading to extra RM uncertainties of about $13.5\,{\rm rad\,m}^{-2}$ for \citetalias{taylor09} sources with $p \lesssim 1$ per cent. This is comparable to the median RM uncertainties of $10.8\,{\rm rad\,m}^{-2}$ reported in the \citetalias{taylor09} catalogue. For a typical \citetalias{taylor09} source, the RM uncertainties should be increased by nine per cent in order to incorporate the effects of off-axis leakage. We further demonstrated that the probability distribution of this extra RM uncertainty is asymmetric, highly non-Gaussian, and is a function of the source RM value. These properties must be carefully taken into account if one wishes to incorporate the off-axis leakage effects to individual sources, which will be the goal of our forthcoming Paper III.

The RM discrepancies of our sources can be largely explained by taking the extra RM uncertainties due to leakage into account, though three sources still show hints of RM time variabilities. Furthermore, by supplementing our dataset with the 282 Smith Cloud sources from \cite{betti19}, we confirmed (at confidence level of about 99.95 per cent) that sources with lower fractional polarisation have larger RM discrepancies between new on-axis VLA observations and that from \citetalias{taylor09}. This is almost certainly due to the residual off-axis polarisation leakage in \citetalias{taylor09} catalogue.

\section*{Acknowledgements}
This is a pre-copyedited, author-produced PDF of an article accepted for publication in the Monthly Notices of the Royal Astronomical Society following peer review. The version of record is available at: xxxxxxx. We thank the anonymous referee for the comments, especially for the suggestion to separate our original manuscript into two stand-alone publications, which have improved the clarity of the papers. We also thank Rainer Beck for his careful reading and valuable suggestions and comments as the MPIfR internal referee, and Aristeidis Noutsos, Shane O'Sullivan, and Dominic Schnitzeler for insightful discussions about this project. Y.K.M.\ was supported for this research by the International Max Planck Research School (IMPRS) for Astronomy and Astrophysics at the Universities of Bonn and Cologne. Y.K.M.\ acknowledges partial support through the Bonn-Cologne Graduate School of Physics and Astronomy. A.B.\ acknowledges financial support by the German Federal Ministry of Education and Research (BMBF) under grant 05A17PB1 (Verbundprojekt D-MeerKAT). A.S.H.\ and S.K.B.\ acknowledge support by NASA through grant number HST-AR-14297 to Haverford College from Space Telescope Science Institute, which is operated by AURA, Inc.\ under NASA contract NAS 5-26555. A.S.H.\ is partially supported by the Dunlap Institute, which is funded through an endowment established by the David Dunlap family and the University of Toronto. The National Radio Astronomy Observatory is a facility of the National Science Foundation operated under cooperative agreement by Associated Universities, Inc. This research has made use of the NASA/IPAC Extragalactic Database (NED) which is operated by the Jet Propulsion Laboratory, California Institute of Technology, under contract with the National Aeronautics and Space Administration. The Wisconsin H-Alpha Mapper and its Sky Survey have been funded primarily through awards from the U.S.\ National Science Foundation.

\bsp

\bibliography{ms}

\begin{thebibliography}{}
\makeatletter
\relax
\def\mn@urlcharsother{\let\do\@makeother \do\$\do\&\do\#\do\^\do\_\do\%\do\~}
\def\mn@doi{\begingroup\mn@urlcharsother \@ifnextchar [ {\mn@doi@}
  {\mn@doi@[]}}
\def\mn@doi@[#1]#2{\def\@tempa{#1}\ifx\@tempa\@empty \href
  {http://dx.doi.org/#2} {doi:#2}\else \href {http://dx.doi.org/#2} {#1}\fi
  \endgroup}
\def\mn@eprint#1#2{\mn@eprint@#1:#2::\@nil}
\def\mn@eprint@arXiv#1{\href {http://arxiv.org/abs/#1} {{\tt arXiv:#1}}}
\def\mn@eprint@dblp#1{\href {http://dblp.uni-trier.de/rec/bibtex/#1.xml}
  {dblp:#1}}
\def\mn@eprint@#1:#2:#3:#4\@nil{\def\@tempa {#1}\def\@tempb {#2}\def\@tempc
  {#3}\ifx \@tempc \@empty \let \@tempc \@tempb \let \@tempb \@tempa \fi \ifx
  \@tempb \@empty \def\@tempb {arXiv}\fi \@ifundefined
  {mn@eprint@\@tempb}{\@tempb:\@tempc}{\expandafter \expandafter \csname
  mn@eprint@\@tempb\endcsname \expandafter{\@tempc}}}

\bibitem[\protect\citeauthoryear{{Abolfathi} et~al.,}{{Abolfathi}
  et~al.}{2018}]{sdssdr14}
{Abolfathi} B.,  et~al., 2018, \mn@doi [\apjs] {10.3847/1538-4365/aa9e8a},
  \href {http://adsabs.harvard.edu/abs/2018ApJS..235...42A} {235, 42}

\bibitem[\protect\citeauthoryear{{Ackermann} et~al.,}{{Ackermann}
  et~al.}{2011}]{fermiagn2}
{Ackermann} M.,  et~al., 2011, \mn@doi [\apj] {10.1088/0004-637X/743/2/171},
  \href {http://adsabs.harvard.edu/abs/2011ApJ...743..171A} {743, 171}

\bibitem[\protect\citeauthoryear{{Adgie}, {Crowther}  \& {Gent}}{{Adgie}
  et~al.}{1972}]{adgie72}
{Adgie} R.~L.,  {Crowther} J.~H.,   {Gent} H.,  1972, \mn@doi [\mnras]
  {10.1093/mnras/159.3.233}, \href
  {http://adsabs.harvard.edu/abs/1972MNRAS.159..233A} {159, 233}

\bibitem[\protect\citeauthoryear{{Aller}}{{Aller}}{1970}]{aller70}
{Aller} H.~D.,  1970, \mn@doi [\apj] {10.1086/150508}, \href
  {http://adsabs.harvard.edu/abs/1970ApJ...161....1A} {161, 1}

\bibitem[\protect\citeauthoryear{{Anderson}, {Gaensler}  \& {Feain}}{{Anderson}
  et~al.}{2016}]{anderson16}
{Anderson} C.~S.,  {Gaensler} B.~M.,   {Feain} I.~J.,  2016, \mn@doi [\apj]
  {10.3847/0004-637X/825/1/59}, \href
  {http://adsabs.harvard.edu/abs/2016ApJ...825...59A} {825, 59}

\bibitem[\protect\citeauthoryear{{Anderson}, {O'Sullivan}, {Heald}, {Hodgson},
  {Pasetto}  \& {Gaensler}}{{Anderson} et~al.}{2019}]{anderson19}
{Anderson} C.~S.,  {O'Sullivan} S.~P.,  {Heald} G.~H.,  {Hodgson} T.,
  {Pasetto} A.,   {Gaensler} B.~M.,  2019, \mn@doi [\mnras]
  {10.1093/mnras/stz377}, \href
  {http://adsabs.harvard.edu/abs/2019MNRAS.485.3600A} {485, 3600}

\bibitem[\protect\citeauthoryear{{Baars}, {Genzel}, {Pauliny-Toth}  \&
  {Witzel}}{{Baars} et~al.}{1977}]{baars77}
{Baars} J.~W.~M.,  {Genzel} R.,  {Pauliny-Toth} I.~I.~K.,   {Witzel} A.,  1977,
  \aap, \href {http://adsabs.harvard.edu/abs/1977A%26A....61...99B} {61, 99}

\bibitem[\protect\citeauthoryear{{Beck}}{{Beck}}{2016}]{beck15}
{Beck} R.,  2016, \mn@doi [\aapr] {10.1007/s00159-015-0084-4}, \href
  {http://adsabs.harvard.edu/abs/2015A%26ARv..24....4B} {24, 4}

\bibitem[\protect\citeauthoryear{{Beck} \& {Wielebinski}}{{Beck} \&
  {Wielebinski}}{2013}]{beck13}
{Beck} R.,  {Wielebinski} R.,  2013, {Magnetic Fields in Galaxies}.
Springer, Berlin, Germany, p.~641

\bibitem[\protect\citeauthoryear{{Becker}, {White}  \& {Edwards}}{{Becker}
  et~al.}{1991}]{becker91}
{Becker} R.~H.,  {White} R.~L.,   {Edwards} A.~L.,  1991, \mn@doi [\apjs]
  {10.1086/191529}, \href {http://adsabs.harvard.edu/abs/1991ApJS...75....1B}
  {75, 1}

\bibitem[\protect\citeauthoryear{{Becker}, {White}  \& {Helfand}}{{Becker}
  et~al.}{1995}]{becker95}
{Becker} R.~H.,  {White} R.~L.,   {Helfand} D.~J.,  1995, \mn@doi [\apj]
  {10.1086/176166}, \href {http://adsabs.harvard.edu/abs/1995ApJ...450..559B}
  {450, 559}

\bibitem[\protect\citeauthoryear{{Betti}, {Hill}, {Mao}, {Gaensler}, {Lockman},
  {McClure-Griffiths}  \& {Benjamin}}{{Betti} et~al.}{2019}]{betti19}
{Betti} S.~K.,  {Hill} A.~S.,  {Mao} S.~A.,  {Gaensler} B.~M.,  {Lockman}
  F.~J.,  {McClure-Griffiths} N.~M.,   {Benjamin} R.~A.,  2019, \mn@doi [\apj]
  {10.3847/1538-4357/aaf886}, \href
  {http://adsabs.harvard.edu/abs/2019ApJ...871..215B} {871, 215}

\bibitem[\protect\citeauthoryear{{Bhatnagar}, {Cornwell}, {Golap}  \&
  {Uson}}{{Bhatnagar} et~al.}{2008}]{bhatnagar08}
{Bhatnagar} S.,  {Cornwell} T.~J.,  {Golap} K.,   {Uson} J.~M.,  2008, \mn@doi
  [\aap] {10.1051/0004-6361:20079284}, \href
  {http://adsabs.harvard.edu/abs/2008A%26A...487..419B} {487, 419}

\bibitem[\protect\citeauthoryear{{Bhatnagar}, {Rau}  \& {Golap}}{{Bhatnagar}
  et~al.}{2013}]{bhatnagar13}
{Bhatnagar} S.,  {Rau} U.,   {Golap} K.,  2013, \mn@doi [\apj]
  {10.1088/0004-637X/770/2/91}, \href
  {http://adsabs.harvard.edu/abs/2013ApJ...770...91B} {770, 91}

\bibitem[\protect\citeauthoryear{{Briggs}}{{Briggs}}{1995}]{briggs95}
{Briggs} D.~S.,  1995, BAAS, \href
  {http://adsabs.harvard.edu/abs/1995AAS...18711202B} {27, 1444}

\bibitem[\protect\citeauthoryear{{Cohen}, {Lane}, {Cotton}, {Kassim}, {Lazio},
  {Perley}, {Condon}  \& {Erickson}}{{Cohen} et~al.}{2007}]{cohen07}
{Cohen} A.~S.,  {Lane} W.~M.,  {Cotton} W.~D.,  {Kassim} N.~E.,  {Lazio}
  T.~J.~W.,  {Perley} R.~A.,  {Condon} J.~J.,   {Erickson} W.~C.,  2007,
  \mn@doi [\aj] {10.1086/520719}, \href
  {http://adsabs.harvard.edu/abs/2007AJ....134.1245C} {134, 1245}

\bibitem[\protect\citeauthoryear{{Condon}, {Cotton}, {Greisen}, {Yin},
  {Perley}, {Taylor}  \& {Broderick}}{{Condon} et~al.}{1998}]{condon98}
{Condon} J.~J.,  {Cotton} W.~D.,  {Greisen} E.~W.,  {Yin} Q.~F.,  {Perley}
  R.~A.,  {Taylor} G.~B.,   {Broderick} J.~J.,  1998, \mn@doi [\aj]
  {10.1086/300337}, \href {http://adsabs.harvard.edu/abs/1998AJ....115.1693C}
  {115, 1693}

\bibitem[\protect\citeauthoryear{{Costa}, {Spangler}, {Sink}, {Brown}  \&
  {Mao}}{{Costa} et~al.}{2016}]{costa16}
{Costa} A.~H.,  {Spangler} S.~R.,  {Sink} J.~R.,  {Brown} S.,   {Mao} S.~A.,
  2016, \mn@doi [\apj] {10.3847/0004-637X/821/2/92}, \href
  {http://adsabs.harvard.edu/abs/2016ApJ...821...92C} {821, 92}

\bibitem[\protect\citeauthoryear{{Cotton}}{{Cotton}}{1994}]{cotton94}
{Cotton} W.~D.,  1994, {Widefield Polarization Correction of VLA Snapshot
  Images at 1.4 GHz}.
NRAO, Charlottesville, VA, USA

\bibitem[\protect\citeauthoryear{{Douglas}, {Bash}, {Bozyan}, {Torrence}  \&
  {Wolfe}}{{Douglas} et~al.}{1996}]{douglas96}
{Douglas} J.~N.,  {Bash} F.~N.,  {Bozyan} F.~A.,  {Torrence} G.~W.,   {Wolfe}
  C.,  1996, \mn@doi [\aj] {10.1086/117932}, \href
  {http://adsabs.harvard.edu/abs/1996AJ....111.1945D} {111, 1945}

\bibitem[\protect\citeauthoryear{{Fletcher}, {Beck}, {Shukurov}, {Berkhuijsen}
  \& {Horellou}}{{Fletcher} et~al.}{2011}]{fletcher11}
{Fletcher} A.,  {Beck} R.,  {Shukurov} A.,  {Berkhuijsen} E.~M.,   {Horellou}
  C.,  2011, \mn@doi [\mnras] {10.1111/j.1365-2966.2010.18065.x}, \href
  {http://adsabs.harvard.edu/abs/2011MNRAS.412.2396F} {412, 2396}

\bibitem[\protect\citeauthoryear{{Gaensler}, {Haverkorn}, {Staveley-Smith},
  {Dickey}, {McClure-Griffiths}, {Dickel}  \& {Wolleben}}{{Gaensler}
  et~al.}{2005}]{gaensler05}
{Gaensler} B.~M.,  {Haverkorn} M.,  {Staveley-Smith} L.,  {Dickey} J.~M.,
  {McClure-Griffiths} N.~M.,  {Dickel} J.~R.,   {Wolleben} M.,  2005, \mn@doi
  [Science] {10.1126/science.1108832}, \href
  {http://adsabs.harvard.edu/abs/2005Sci...307.1610G} {307, 1610}

\bibitem[\protect\citeauthoryear{{Gaensler}, {Landecker}, {Taylor}  \& {POSSUM
  Collaboration}}{{Gaensler} et~al.}{2010}]{gaensler10}
{Gaensler} B.~M.,  {Landecker} T.~L.,  {Taylor} A.~R.,   {POSSUM Collaboration}
  2010, BAAS, \href {http://adsabs.harvard.edu/abs/2010AAS...21547013G} {42,
  515}

\bibitem[\protect\citeauthoryear{{Gie{\ss}{\"u}bel}, {Heald}, {Beck}  \&
  {Arshakian}}{{Gie{\ss}{\"u}bel} et~al.}{2013}]{giessuebel13}
{Gie{\ss}{\"u}bel} R.,  {Heald} G.,  {Beck} R.,   {Arshakian} T.~G.,  2013,
  \mn@doi [\aap] {10.1051/0004-6361/201321765}, \href
  {http://adsabs.harvard.edu/abs/2013A%26A...559A..27G} {559, A27}

\bibitem[\protect\citeauthoryear{{Gower}, {Scott}  \& {Wills}}{{Gower}
  et~al.}{1967}]{gower67}
{Gower} J.~F.~R.,  {Scott} P.~F.,   {Wills} D.,  1967, \memras, \href
  {http://adsabs.harvard.edu/abs/1967MmRAS..71...49G} {71, 49}

\bibitem[\protect\citeauthoryear{{Gregory} \& {Condon}}{{Gregory} \&
  {Condon}}{1991}]{gregory91}
{Gregory} P.~C.,  {Condon} J.~J.,  1991, \mn@doi [\apjs] {10.1086/191559},
  \href {http://adsabs.harvard.edu/abs/1991ApJS...75.1011G} {75, 1011}

\bibitem[\protect\citeauthoryear{{Griffith}, {Wright}, {Burke}  \&
  {Ekers}}{{Griffith} et~al.}{1994}]{griffith94}
{Griffith} M.~R.,  {Wright} A.~E.,  {Burke} B.~F.,   {Ekers} R.~D.,  1994,
  \mn@doi [\apjs] {10.1086/191863}, \href
  {http://adsabs.harvard.edu/abs/1994ApJS...90..179G} {90, 179}

\bibitem[\protect\citeauthoryear{{Hales}}{{Hales}}{2017}]{hales17}
{Hales} C.~A.,  2017, \mn@doi [\aj] {10.3847/1538-3881/aa7aef}, \href
  {http://adsabs.harvard.edu/abs/2017AJ....154...54H} {154, 54}

\bibitem[\protect\citeauthoryear{{Harvey-Smith}, {Madsen}  \&
  {Gaensler}}{{Harvey-Smith} et~al.}{2011}]{harvey-smith11}
{Harvey-Smith} L.,  {Madsen} G.~J.,   {Gaensler} B.~M.,  2011, \mn@doi [\apj]
  {10.1088/0004-637X/736/2/83}, \href
  {http://adsabs.harvard.edu/abs/2011ApJ...736...83H} {736, 83}

\bibitem[\protect\citeauthoryear{{Hewitt} \& {Burbidge}}{{Hewitt} \&
  {Burbidge}}{1987}]{hewitt87}
{Hewitt} A.,  {Burbidge} G.,  1987, \mn@doi [\apjs] {10.1086/191163}, \href
  {http://adsabs.harvard.edu/abs/1987ApJS...63....1H} {63, 1}

\bibitem[\protect\citeauthoryear{{Hewitt} \& {Burbidge}}{{Hewitt} \&
  {Burbidge}}{1991}]{hewitt91}
{Hewitt} A.,  {Burbidge} G.,  1991, \mn@doi [\apjs] {10.1086/191533}, \href
  {http://adsabs.harvard.edu/abs/1991ApJS...75..297H} {75, 297}

\bibitem[\protect\citeauthoryear{{Hill}, {Mao}, {Benjamin}, {Lockman}  \&
  {McClure-Griffiths}}{{Hill} et~al.}{2013}]{hill13}
{Hill} A.~S.,  {Mao} S.~A.,  {Benjamin} R.~A.,  {Lockman} F.~J.,
  {McClure-Griffiths} N.~M.,  2013, \mn@doi [\apj]
  {10.1088/0004-637X/777/1/55}, \href
  {http://adsabs.harvard.edu/abs/2013ApJ...777...55H} {777, 55}

\bibitem[\protect\citeauthoryear{{Hovatta}, {Lister}, {Aller}, {Aller},
  {Homan}, {Kovalev}, {Pushkarev}  \& {Savolainen}}{{Hovatta}
  et~al.}{2012}]{hovatta12}
{Hovatta} T.,  {Lister} M.~L.,  {Aller} M.~F.,  {Aller} H.~D.,  {Homan} D.~C.,
  {Kovalev} Y.~Y.,  {Pushkarev} A.~B.,   {Savolainen} T.,  2012, \mn@doi [\aj]
  {10.1088/0004-6256/144/4/105}, \href
  {http://adsabs.harvard.edu/abs/2012AJ....144..105H} {144, 105}

\bibitem[\protect\citeauthoryear{{Huchra} et~al.,}{{Huchra}
  et~al.}{2012}]{2mass}
{Huchra} J.~P.,  et~al., 2012, \mn@doi [\apjs] {10.1088/0067-0049/199/2/26},
  \href {http://adsabs.harvard.edu/abs/2012ApJS..199...26H} {199, 26}

\bibitem[\protect\citeauthoryear{{Jagannathan}, {Bhatnagar}, {Rau}  \&
  {Taylor}}{{Jagannathan} et~al.}{2017}]{jagannathan17}
{Jagannathan} P.,  {Bhatnagar} S.,  {Rau} U.,   {Taylor} A.~R.,  2017, \mn@doi
  [\aj] {10.3847/1538-3881/aa77f8}, \href
  {http://adsabs.harvard.edu/abs/2017AJ....154...56J} {154, 56}

\bibitem[\protect\citeauthoryear{{Jagannathan}, {Bhatnagar}, {Brisken}  \&
  {Taylor}}{{Jagannathan} et~al.}{2018}]{jagannathan18}
{Jagannathan} P.,  {Bhatnagar} S.,  {Brisken} W.,   {Taylor} A.~R.,  2018,
  \mn@doi [\aj] {10.3847/1538-3881/aa989f}, \href
  {http://adsabs.harvard.edu/abs/2018AJ....155....3J} {155, 3}

\bibitem[\protect\citeauthoryear{{Jones} et~al.,}{{Jones} et~al.}{2009}]{6dFGS}
{Jones} D.~H.,  et~al., 2009, \mn@doi [\mnras]
  {10.1111/j.1365-2966.2009.15338.x}, \href
  {http://adsabs.harvard.edu/abs/2009MNRAS.399..683J} {399, 683}

\bibitem[\protect\citeauthoryear{{Kaczmarek}, {Purcell}, {Gaensler},
  {McClure-Griffiths}  \& {Stevens}}{{Kaczmarek} et~al.}{2017}]{kaczmarek17}
{Kaczmarek} J.~F.,  {Purcell} C.~R.,  {Gaensler} B.~M.,  {McClure-Griffiths}
  N.~M.,   {Stevens} J.,  2017, \mn@doi [\mnras] {10.1093/mnras/stx206}, \href
  {http://adsabs.harvard.edu/abs/2017MNRAS.467.1776K} {467, 1776}

\bibitem[\protect\citeauthoryear{{Kierdorf}, {Beck}, {Hoeft}, {Klein}, {van
  Weeren}, {Forman}  \& {Jones}}{{Kierdorf} et~al.}{2017}]{kierdorf17}
{Kierdorf} M.,  {Beck} R.,  {Hoeft} M.,  {Klein} U.,  {van Weeren} R.~J.,
  {Forman} W.~R.,   {Jones} C.,  2017, \mn@doi [\aap]
  {10.1051/0004-6361/201629570}, \href
  {http://adsabs.harvard.edu/abs/2017A%26A...600A..18K} {600, A18}

\bibitem[\protect\citeauthoryear{{Kuehr}, {Witzel}, {Pauliny-Toth}  \&
  {Nauber}}{{Kuehr} et~al.}{1981}]{kuehr81}
{Kuehr} H.,  {Witzel} A.,  {Pauliny-Toth} I.~I.~K.,   {Nauber} U.,  1981,
  \aaps, \href {http://adsabs.harvard.edu/abs/1981A%26AS...45..367K} {45, 367}

\bibitem[\protect\citeauthoryear{{Large}, {Mills}, {Little}, {Crawford}  \&
  {Sutton}}{{Large} et~al.}{1981}]{large81}
{Large} M.~I.,  {Mills} B.~Y.,  {Little} A.~G.,  {Crawford} D.~F.,   {Sutton}
  J.~M.,  1981, \mn@doi [\mnras] {10.1093/mnras/194.3.693}, \href
  {http://adsabs.harvard.edu/abs/1981MNRAS.194..693L} {194, 693}

\bibitem[\protect\citeauthoryear{{Ma}, {Mao}, {Stil}, {Basu}, {West}, {Heiles},
  {Hill}  \& {Betti}}{{Ma} et~al.}{2019}]{ma19a}
{Ma} Y.~K.,  {Mao} S.~A.,  {Stil} J.,  {Basu} A.,  {West} J.,  {Heiles} C.,
  {Hill} A.~S.,   {Betti} S.~K.,  2019, \mnras, accepted

\bibitem[\protect\citeauthoryear{{Mao}, {Gaensler}, {Stanimirovi{\'c}},
  {Haverkorn}, {McClure-Griffiths}, {Staveley-Smith}  \& {Dickey}}{{Mao}
  et~al.}{2008}]{mao08}
{Mao} S.~A.,  {Gaensler} B.~M.,  {Stanimirovi{\'c}} S.,  {Haverkorn} M.,
  {McClure-Griffiths} N.~M.,  {Staveley-Smith} L.,   {Dickey} J.~M.,  2008,
  \mn@doi [\apj] {10.1086/590546}, \href
  {http://adsabs.harvard.edu/abs/2008ApJ...688.1029M} {688, 1029}

\bibitem[\protect\citeauthoryear{{Mao}, {Gaensler}, {Haverkorn}, {Zweibel},
  {Madsen}, {McClure-Griffiths}, {Shukurov}  \& {Kronberg}}{{Mao}
  et~al.}{2010}]{mao10}
{Mao} S.~A.,  {Gaensler} B.~M.,  {Haverkorn} M.,  {Zweibel} E.~G.,  {Madsen}
  G.~J.,  {McClure-Griffiths} N.~M.,  {Shukurov} A.,   {Kronberg} P.~P.,  2010,
  \mn@doi [\apj] {10.1088/0004-637X/714/2/1170}, \href
  {http://adsabs.harvard.edu/abs/2010ApJ...714.1170M} {714, 1170}

\bibitem[\protect\citeauthoryear{{Mao} et~al.,}{{Mao} et~al.}{2012}]{mao12}
{Mao} S.~A.,  et~al., 2012, \mn@doi [\apj] {10.1088/0004-637X/755/1/21}, \href
  {http://adsabs.harvard.edu/abs/2012ApJ...755...21M} {755, 21}

\bibitem[\protect\citeauthoryear{{Mao}, {Zweibel}, {Fletcher}, {Ott}  \&
  {Tabatabaei}}{{Mao} et~al.}{2015}]{mao15}
{Mao} S.~A.,  {Zweibel} E.,  {Fletcher} A.,  {Ott} J.,   {Tabatabaei} F.,
  2015, \mn@doi [\apj] {10.1088/0004-637X/800/2/92}, \href
  {http://adsabs.harvard.edu/abs/2015ApJ...800...92M} {800, 92}

\bibitem[\protect\citeauthoryear{{Mao} et~al.,}{{Mao} et~al.}{2017}]{mao17}
{Mao} S.~A.,  et~al., 2017, \mn@doi [Nature Astronomy]
  {10.1038/s41550-017-0218-x}, \href
  {http://adsabs.harvard.edu/abs/2017NatAs...1..621M} {1, 621}

\bibitem[\protect\citeauthoryear{{Massardi} et~al.,}{{Massardi}
  et~al.}{2008}]{massardi08}
{Massardi} M.,  et~al., 2008, \mn@doi [\mnras]
  {10.1111/j.1365-2966.2007.12751.x}, \href
  {http://adsabs.harvard.edu/abs/2008MNRAS.384..775M} {384, 775}

\bibitem[\protect\citeauthoryear{{Mauch}, {Murphy}, {Buttery}, {Curran},
  {Hunstead}, {Piestrzynski}, {Robertson}  \& {Sadler}}{{Mauch}
  et~al.}{2003}]{mauch03}
{Mauch} T.,  {Murphy} T.,  {Buttery} H.~J.,  {Curran} J.,  {Hunstead} R.~W.,
  {Piestrzynski} B.,  {Robertson} J.~G.,   {Sadler} E.~M.,  2003, \mn@doi
  [\mnras] {10.1046/j.1365-8711.2003.06605.x}, \href
  {http://adsabs.harvard.edu/abs/2003MNRAS.342.1117M} {342, 1117}

\bibitem[\protect\citeauthoryear{{McClure-Griffiths}, {Madsen}, {Gaensler},
  {McConnell}  \& {Schnitzeler}}{{McClure-Griffiths}
  et~al.}{2010}]{mcclure-griffiths10}
{McClure-Griffiths} N.~M.,  {Madsen} G.~J.,  {Gaensler} B.~M.,  {McConnell} D.,
    {Schnitzeler} D.~H.~F.~M.,  2010, \mn@doi [\apj]
  {10.1088/0004-637X/725/1/275}, \href
  {http://adsabs.harvard.edu/abs/2010ApJ...725..275M} {725, 275}

\bibitem[\protect\citeauthoryear{{McMullin}, {Waters}, {Schiebel}, {Young}  \&
  {Golap}}{{McMullin} et~al.}{2007}]{mcmullin07}
{McMullin} J.~P.,  {Waters} B.,  {Schiebel} D.,  {Young} W.,   {Golap} K.,
  2007, in {Shaw} R.~A.,  {Hill} F.,   {Bell} D.~J.,  eds,  ASP Conf. Ser. Vol.
  376, Astronomical Data Analysis Software and Systems XVI. ASP, San Francisco,
  CA, USA, p.~127

\bibitem[\protect\citeauthoryear{{Murphy} et~al.,}{{Murphy}
  et~al.}{2010}]{murphy10}
{Murphy} T.,  et~al., 2010, \mn@doi [\mnras]
  {10.1111/j.1365-2966.2009.15961.x}, \href
  {http://adsabs.harvard.edu/abs/2010MNRAS.402.2403M} {402, 2403}

\bibitem[\protect\citeauthoryear{{Myers}, {Baum}  \& {Chandler}}{{Myers}
  et~al.}{2014}]{myers14}
{Myers} S.~T.,  {Baum} S.~A.,   {Chandler} C.~J.,  2014, BAAS, \href
  {http://adsabs.harvard.edu/abs/2014AAS...22323601M} {223, 236.01}

\bibitem[\protect\citeauthoryear{{Oppermann} et~al.,}{{Oppermann}
  et~al.}{2012}]{oppermann12}
{Oppermann} N.,  et~al., 2012, \mn@doi [\aap] {10.1051/0004-6361/201118526},
  \href {http://adsabs.harvard.edu/abs/2012A%26A...542A..93O} {542, A93}

\bibitem[\protect\citeauthoryear{{Oppermann} et~al.,}{{Oppermann}
  et~al.}{2015}]{oppermann15}
{Oppermann} N.,  et~al., 2015, \mn@doi [\aap] {10.1051/0004-6361/201423995},
  \href {http://adsabs.harvard.edu/abs/2015A%26A...575A.118O} {575, A118}

\bibitem[\protect\citeauthoryear{{Perley} \& {Butler}}{{Perley} \&
  {Butler}}{2013}]{perley13a}
{Perley} R.~A.,  {Butler} B.~J.,  2013, \mn@doi [\apjs]
  {10.1088/0067-0049/204/2/19}, \href
  {http://adsabs.harvard.edu/abs/2013ApJS..204...19P} {204, 19}

\bibitem[\protect\citeauthoryear{{Perlman}, {Padovani}, {Giommi}, {Sambruna},
  {Jones}, {Tzioumis}  \& {Reynolds}}{{Perlman} et~al.}{1998}]{cxrb}
{Perlman} E.~S.,  {Padovani} P.,  {Giommi} P.,  {Sambruna} R.,  {Jones} L.~R.,
  {Tzioumis} A.,   {Reynolds} J.,  1998, \mn@doi [\aj] {10.1086/300283}, \href
  {http://adsabs.harvard.edu/abs/1998AJ....115.1253P} {115, 1253}

\bibitem[\protect\citeauthoryear{{Purcell} et~al.,}{{Purcell}
  et~al.}{2015}]{purcell15}
{Purcell} C.~R.,  et~al., 2015, \mn@doi [\apj] {10.1088/0004-637X/804/1/22},
  \href {http://adsabs.harvard.edu/abs/2015ApJ...804...22P} {804, 22}

\bibitem[\protect\citeauthoryear{{Richards} et~al.,}{{Richards}
  et~al.}{2011}]{richards11}
{Richards} J.~L.,  et~al., 2011, \mn@doi [\apjs] {10.1088/0067-0049/194/2/29},
  \href {http://adsabs.harvard.edu/abs/2011ApJS..194...29R} {194, 29}

\bibitem[\protect\citeauthoryear{{Rudnick} et~al.,}{{Rudnick}
  et~al.}{1985}]{rudnick85}
{Rudnick} L.,  et~al., 1985, \mn@doi [\apjs] {10.1086/191023}, \href
  {http://adsabs.harvard.edu/abs/1985ApJS...57..693R} {57, 693}

\bibitem[\protect\citeauthoryear{{Shimmins} \& {Wall}}{{Shimmins} \&
  {Wall}}{1973}]{shimmins73}
{Shimmins} A.~J.,  {Wall} J.~V.,  1973, \mn@doi [Australian Journal of Physics]
  {10.1071/PH730093}, \href {http://adsabs.harvard.edu/abs/1973AuJPh..26...93S}
  {26, 93}

\bibitem[\protect\citeauthoryear{{Stil}, {Taylor}  \& {Sunstrum}}{{Stil}
  et~al.}{2011}]{stil11}
{Stil} J.~M.,  {Taylor} A.~R.,   {Sunstrum} C.,  2011, \mn@doi [\apj]
  {10.1088/0004-637X/726/1/4}, \href
  {http://adsabs.harvard.edu/abs/2011ApJ...726....4S} {726, 4}

\bibitem[\protect\citeauthoryear{{Taylor}, {Stil}  \& {Sunstrum}}{{Taylor}
  et~al.}{2009}]{taylor09}
{Taylor} A.~R.,  {Stil} J.~M.,   {Sunstrum} C.,  2009, \mn@doi [\apj]
  {10.1088/0004-637X/702/2/1230}, \href
  {http://adsabs.harvard.edu/abs/2009ApJ...702.1230T} {702, 1230}

\bibitem[\protect\citeauthoryear{{Terral} \& {Ferri{\`e}re}}{{Terral} \&
  {Ferri{\`e}re}}{2017}]{terral17}
{Terral} P.,  {Ferri{\`e}re} K.,  2017, \mn@doi [\aap]
  {10.1051/0004-6361/201629572}, \href
  {http://adsabs.harvard.edu/abs/2017A%26A...600A..29T} {600, A29}

\bibitem[\protect\citeauthoryear{{Van Eck} et~al.,}{{Van Eck}
  et~al.}{2011}]{vaneck11}
{Van Eck} C.~L.,  et~al., 2011, \mn@doi [\apj] {10.1088/0004-637X/728/2/97},
  \href {http://adsabs.harvard.edu/abs/2011ApJ...728...97V} {728, 97}

\bibitem[\protect\citeauthoryear{{Wills}}{{Wills}}{1975}]{wills75}
{Wills} B.~J.,  1975, Australian Journal of Physics Astrophysical Supplement,
  \href {http://adsabs.harvard.edu/abs/1975AuJPA..38....1W} {38, 1}

\bibitem[\protect\citeauthoryear{{Wright}, {Griffith}, {Hunt}, {Troup}, {Burke}
   \& {Ekers}}{{Wright} et~al.}{1996}]{wright96}
{Wright} A.~E.,  {Griffith} M.~R.,  {Hunt} A.~J.,  {Troup} E.,  {Burke} B.~F.,
   {Ekers} R.~D.,  1996, \mn@doi [\apjs] {10.1086/192272}, \href
  {http://adsabs.harvard.edu/abs/1996ApJS..103..145W} {103, 145}

\bibitem[\protect\citeauthoryear{{Zavala} \& {Taylor}}{{Zavala} \&
  {Taylor}}{2001}]{zavala01}
{Zavala} R.~T.,  {Taylor} G.~B.,  2001, \mn@doi [\apjl] {10.1086/319653}, \href
  {http://adsabs.harvard.edu/abs/2001ApJ...550L.147Z} {550, L147}

\makeatother
\end{thebibliography}

\appendix

\section{Radio Spectra of Our Targets}

\begin{figure*}
\includegraphics[width=\doublecolumnwidth]{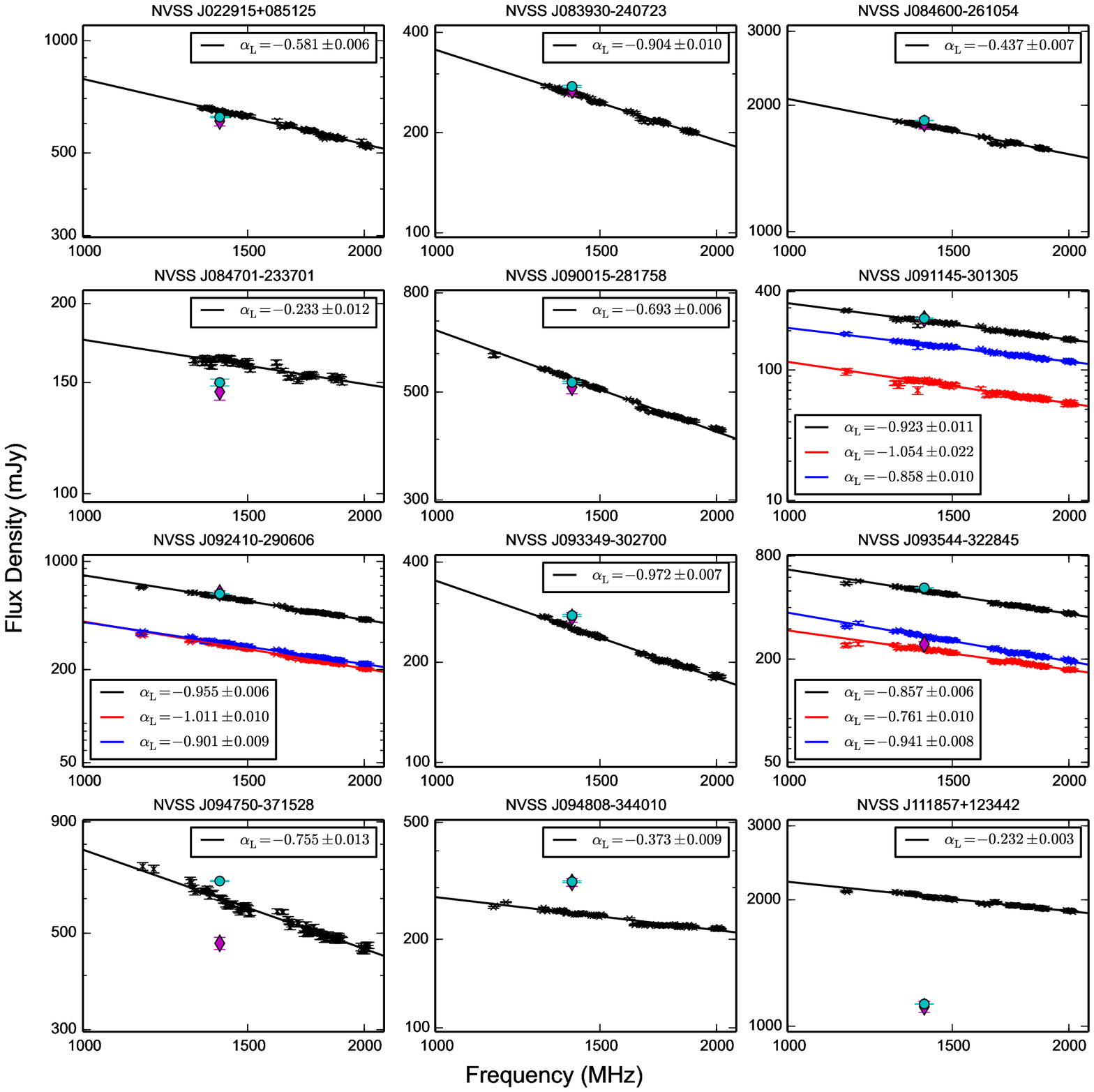}
\caption{Radio spectra across L-band of our target sources from the new VLA observations. The total flux densities are represented by black data points. For sources that are resolved into two spatial components, the flux densities of the individual components are plotted in red (component a) and blue (component b). The best-fit power law spectra ($S_\nu \propto \nu^{\alpha_{\rm L}}$) are shown as solid lines with corresponding colours to the data points. The flux densities we determined from NVSS cutout images ($S_{\rm cutout}$) are plotted as the cyan circles, while that listed in the NVSS catalogue \citep[$S_{\rm NVSS}$;][]{condon98} are plotted as the magenta diamonds.\label{fig:stokesi}}
\end{figure*}

\begin{figure*}
\ContinuedFloat
\includegraphics[width=\doublecolumnwidth]{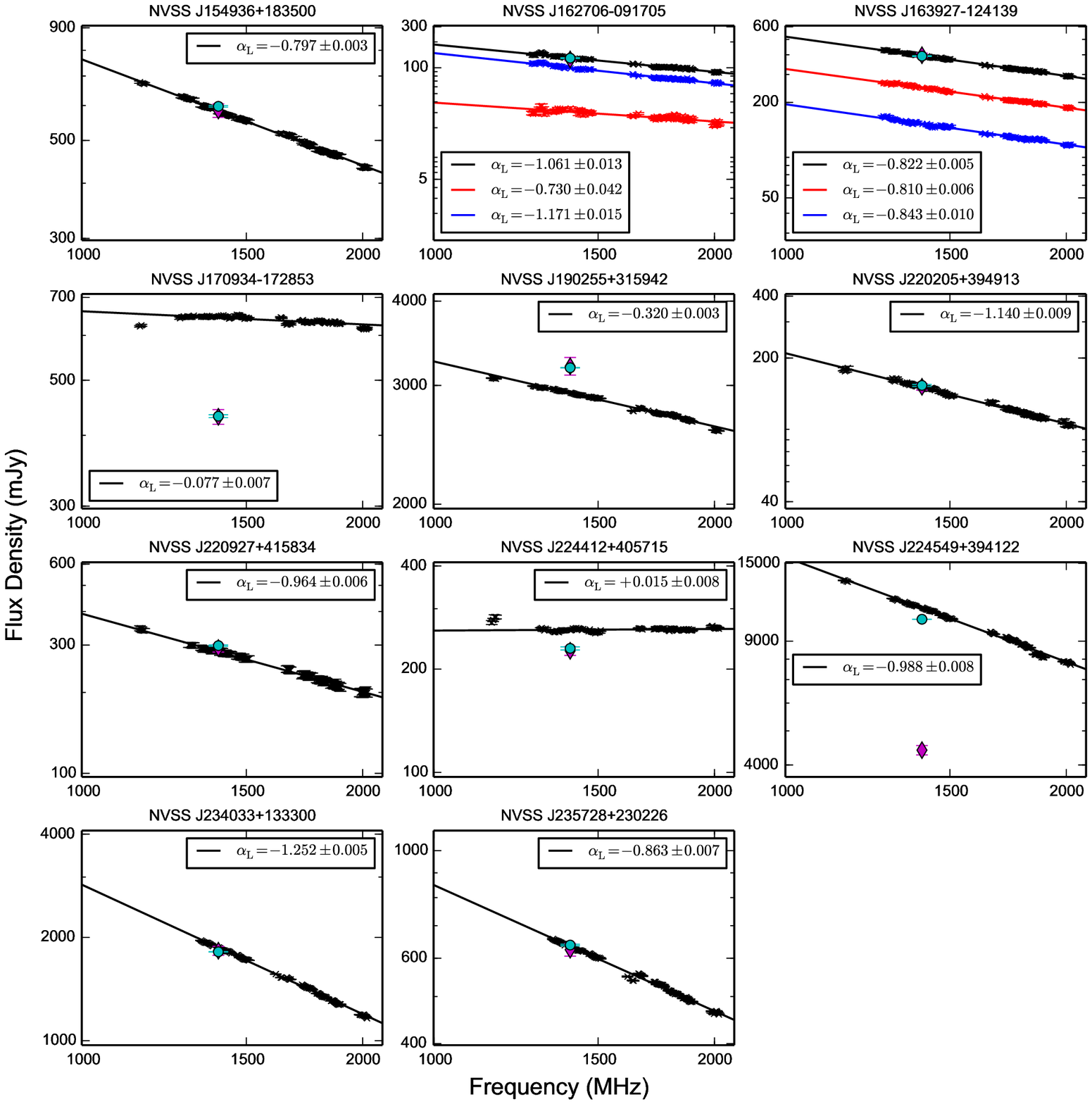}
\caption{(Continued) Radio spectra of our target sources from new VLA observations.}
\end{figure*}

We show the radio spectra of our target sources in Figure~\ref{fig:stokesi}. A simple power law is fitted to our broadband VLA data (from Paper I). For sources identified as spatial doubles, we fitted each component individually, as well as their sum (with uncertainties added in quadrature). The best-fit parameters are listed in Table~\ref{table:stokesi}.

\section{The Role of $\mathbf{PA}_\mathbf{leak}$} \label{sec:paleak_sim}
\begin{figure*}
\includegraphics[width=160.0pt]{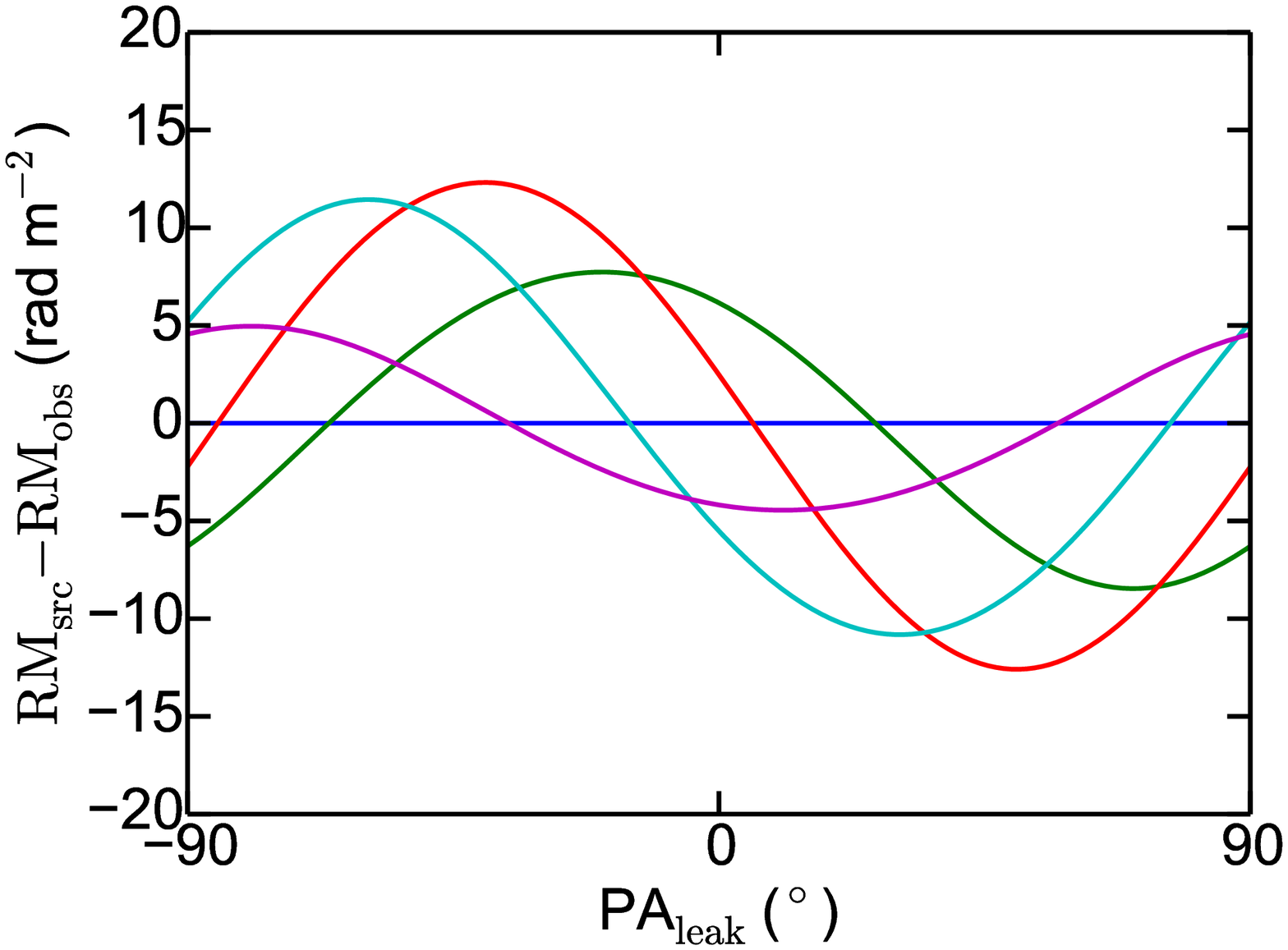}
\includegraphics[width=160.0pt]{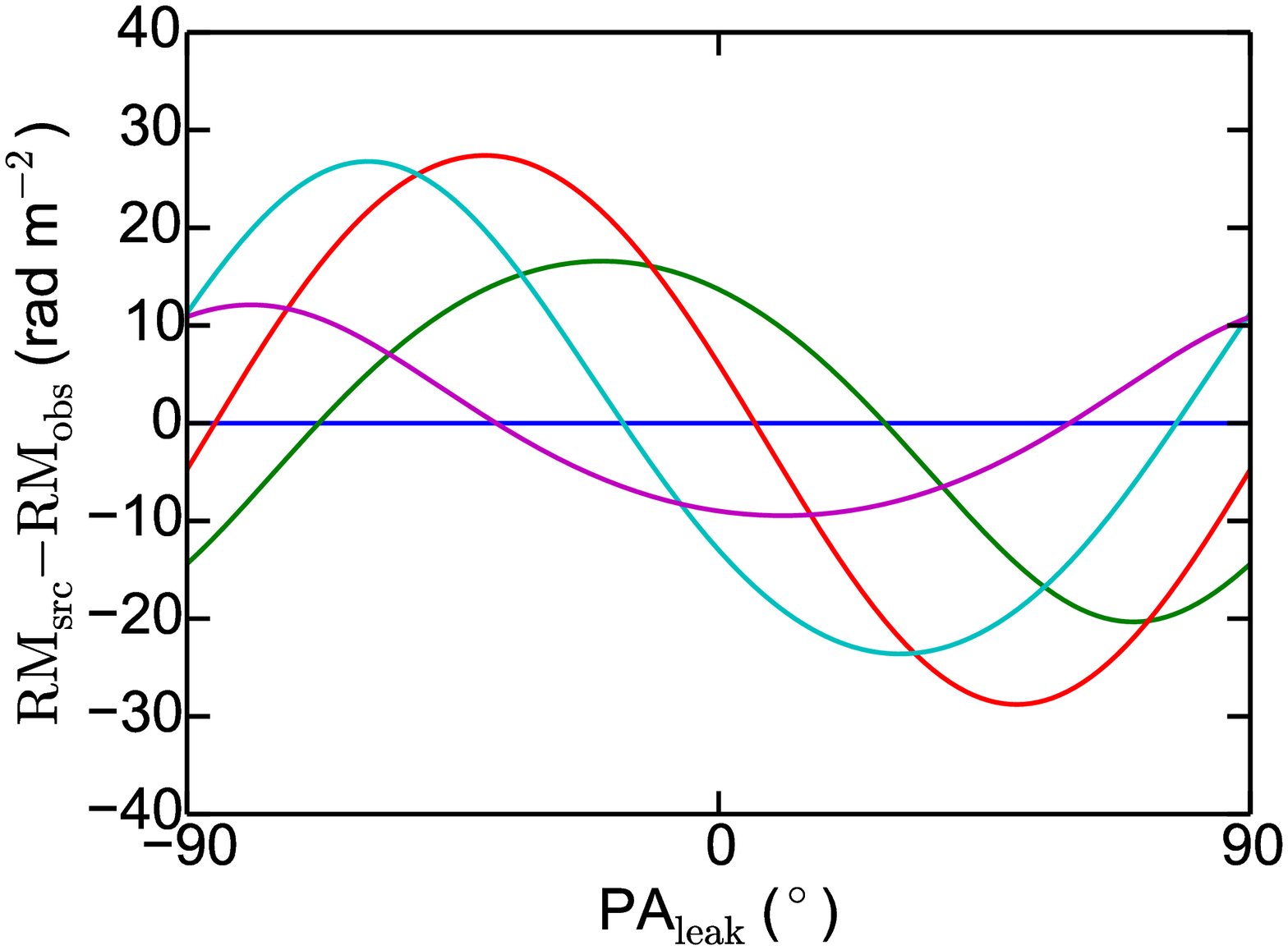}
\includegraphics[width=160.0pt]{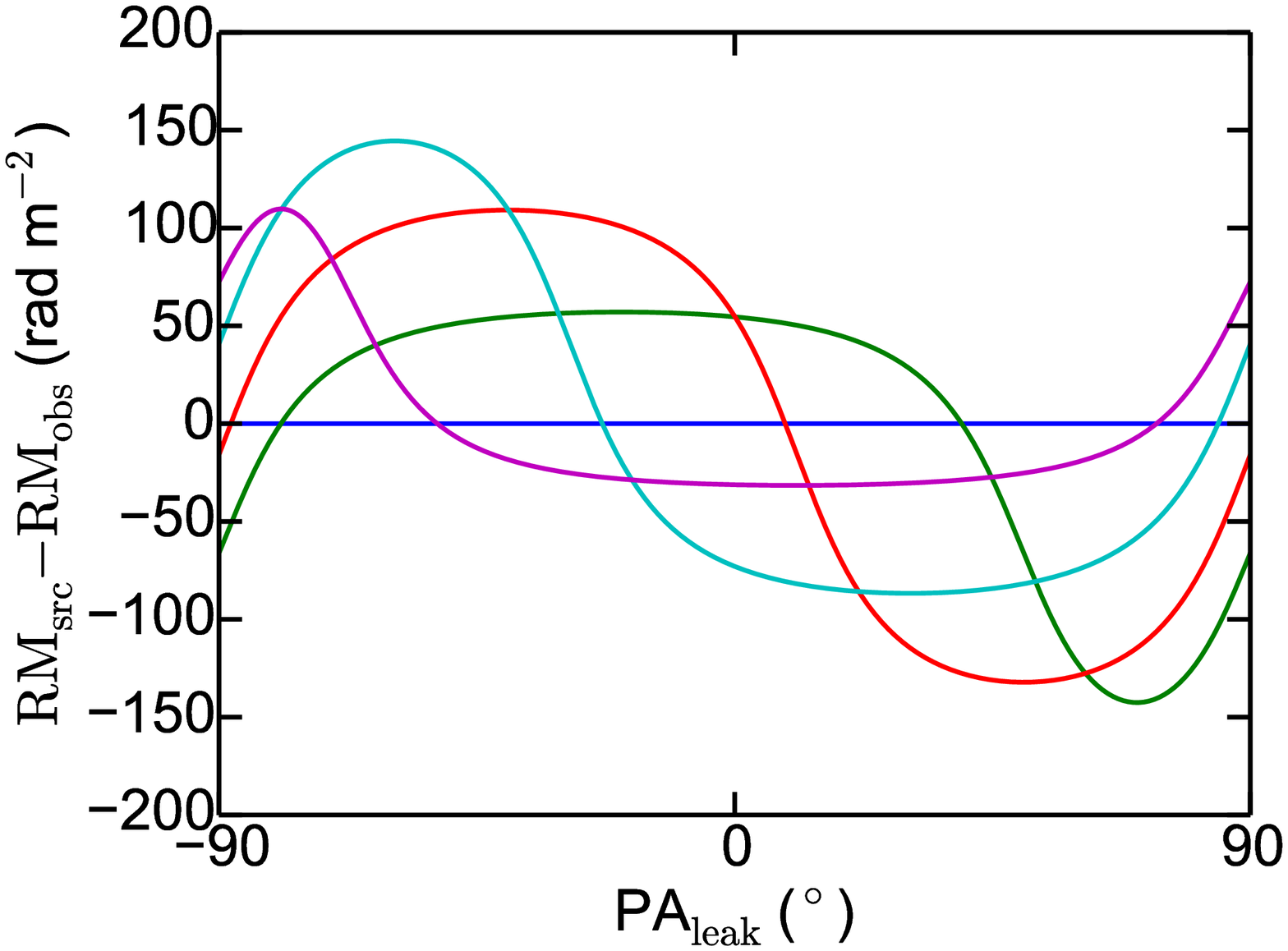}
\caption{Simulation results showing the relationship between $({\rm RM}_{\rm src} - {\rm RM}_{\rm obs})$ and ${\rm PA}_{\rm leak}$. The left, middle, and right panels show the cases where the artificial target sources are strongly ($p = 8.5$ per cent), intermediately ($p = 3.8$ per cent), and weakly ($p = 0.8$ per cent) polarised, respectively. The artificial sources with ${\rm RM}_{\rm in}$ of $0$, $+150$, $+300$, $+450$, and $+600\,{\rm rad\,m}^{-2}$ are shown as the blue, green, red, cyan, and magenta lines, respectively. The input ${\rm PA}_0$ has been chosen such that the true PA at the NVSS IF1 is $0^\circ$. Note that the $y$-axis scales are different among the panels. \label{fig:paleakrm}}
\end{figure*}

\begin{figure*}
\includegraphics[width=160.0pt]{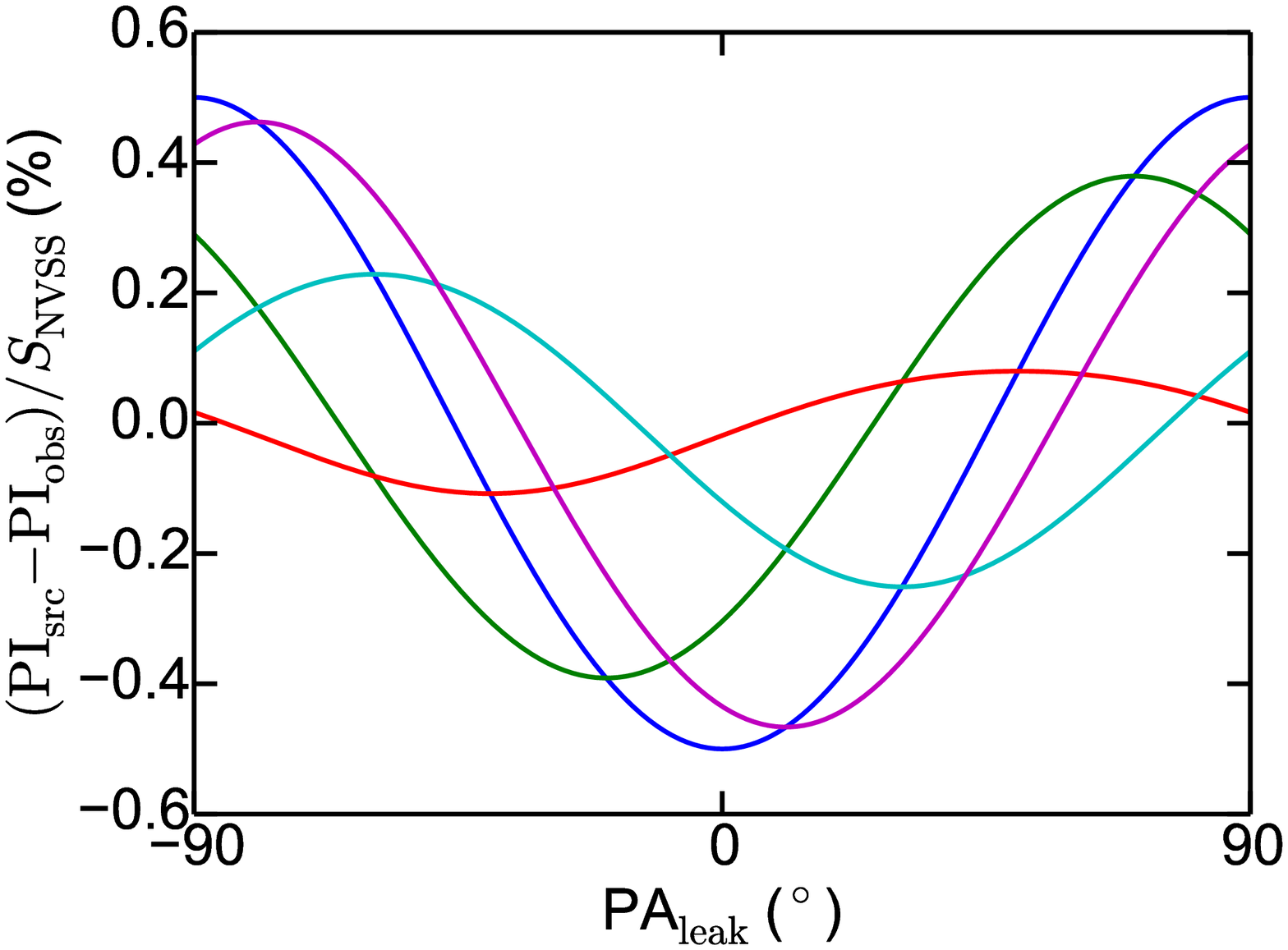}
\includegraphics[width=160.0pt]{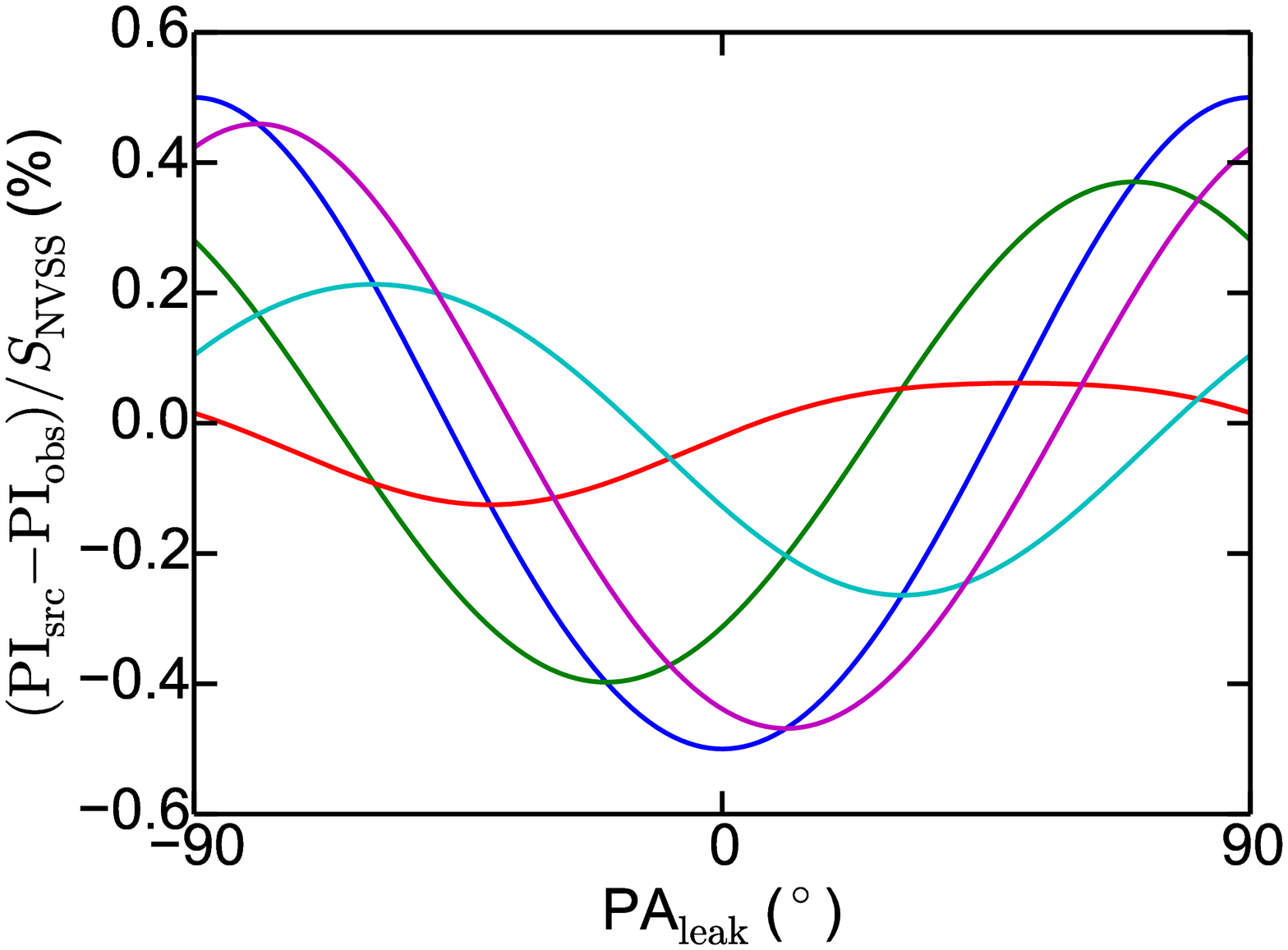}
\includegraphics[width=160.0pt]{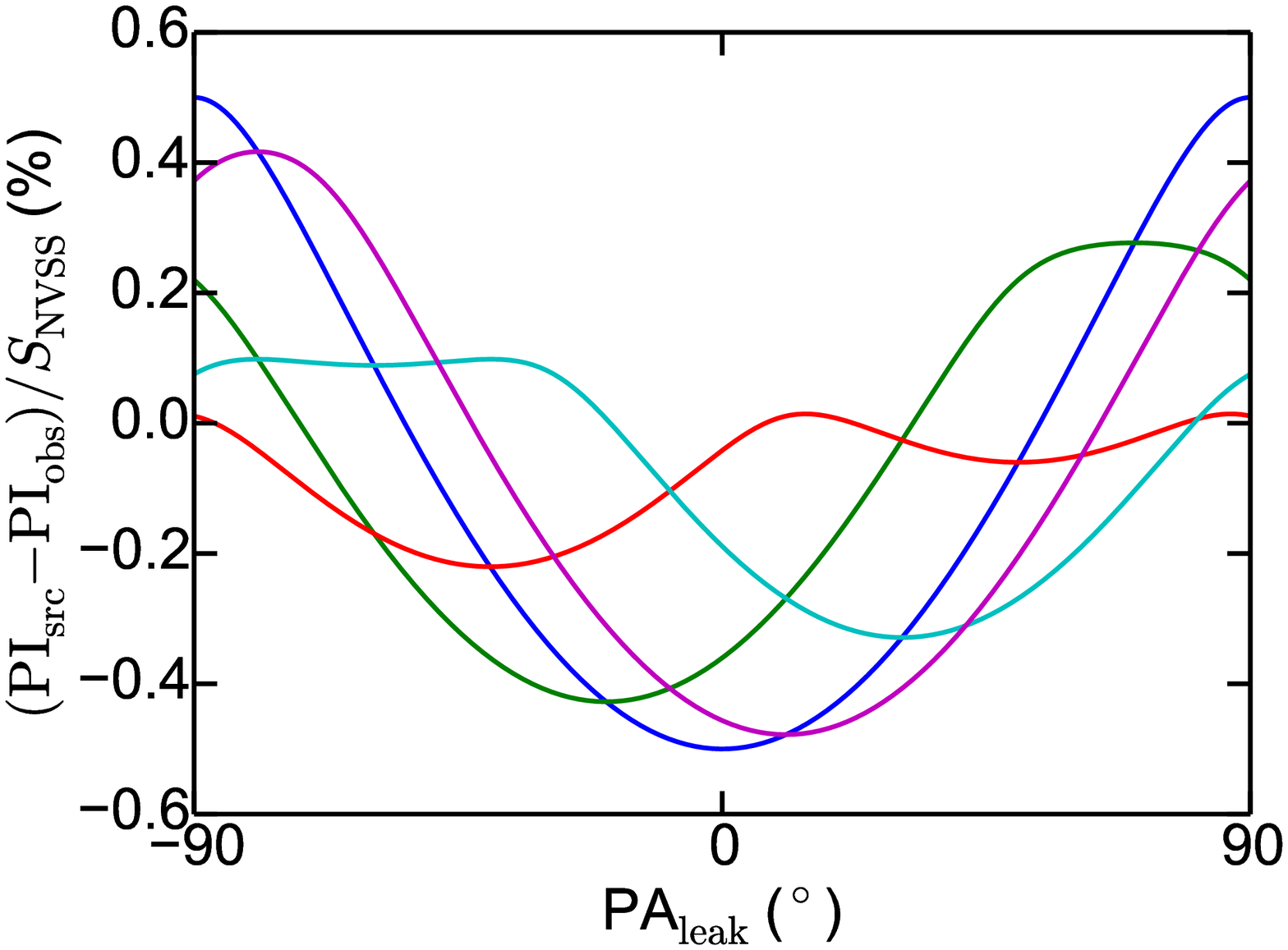}
\caption{Simulation results similar to those in Figure~\ref{fig:paleakrm}, but showing $({\rm PI}_{\rm src} - {\rm PI}_{\rm obs})/S_{\rm NVSS}$ instead as the $y$-axis. \label{fig:paleakpf}}
\label{lastpage}
\end{figure*}

We further investigate here the effects of off-axis polarisation leakage in Section~\ref{sec:leakfrm}. Specifically, we aim to understand the cause of the asymmetric double-horn $({\rm RM}_{\rm src} - {\rm RM}_{\rm obs})$ distributions seen in Figures~\ref{fig:2dhist} and \ref{fig:1dhist}. To achieve this, we repeated our simulation with the artificial sources, again with the \citetalias{taylor09} source properties of J091145$-$301305 ($S_{\rm NVSS} = 247.1\,{\rm mJy}$ and ${\rm PI} = 21.1\,{\rm mJy}$, as the strongly polarised case), J093349$-$302700 ($S_{\rm NVSS} = 272.9\,{\rm mJy}$ and ${\rm PI} = 10.3\,{\rm mJy}$, as the intermediately polarised case), and J235728$+$230226 ($S_{\rm NVSS} = 624.6\,{\rm mJy}$ and ${\rm PI} = 5.0\,{\rm mJy}$, as the weakly polarised case). For each case, we manually selected ${\rm RM}_{\rm src}$ values of $0$, $+150$, $+300$, $+450$, and $+600\,{\rm rad\,m}^{-2}$. However, instead of randomising ${\rm PA}_0$ and ${\rm PA}_{\rm leak}$ as we did in Section~\ref{sec:leakfrm}, here we chose ${\rm PA}_0$ for each artificial source such that the source PA in the NVSS IF1 is $0^\circ$, and ${\rm PA}_{\rm leak}$ is uniformly sampled within $[-\pi/2, +\pi/2]$ to see its effect on $({\rm RM}_{\rm src} - {\rm RM}_{\rm obs})$. The results of this are presented in Figures~\ref{fig:paleakrm} and \ref{fig:paleakpf}, with the former showing the trend of $({\rm RM}_{\rm src} - {\rm RM}_{\rm obs})$ and the latter showing that of $({\rm PI}_{\rm src} - {\rm PI}_{\rm obs})/S_{\rm NVSS}$. By consulting these Figures, we can pinpoint the situations (in terms of the relative PA between the source and leakage vectors) that resulted in the double horn.

From Figure~\ref{fig:paleakrm}, we can see that $({\rm RM}_{\rm src} - {\rm RM}_{\rm obs})$ shows a nearly sinusoidal variation across ${\rm PA}_{\rm leak}$, with the peaks and troughs corresponding to the two horns in Figures~\ref{fig:2dhist} and \ref{fig:1dhist}. For the strongly and intermediately polarised cases, the widths of the peaks and troughs are very similar, leading to the symmetric $({\rm RM}_{\rm src} - {\rm RM}_{\rm obs})$ distributions. On the other hand, for the weakly polarised case the differences in the widths of the peaks compared to that of the troughs are much more apparent. This in turn leads to the extreme asymmetry in the double horns -- the wider (or narrower) part in $({\rm RM}_{\rm src} - {\rm RM}_{\rm obs})$ against ${\rm PA}_{\rm leak}$ (Figure~\ref{fig:paleakrm}) corresponds to the taller (or shorter) of the double horn of the histogram in Figures~\ref{fig:2dhist} and \ref{fig:1dhist}.

\end{document}